\documentclass[aps,prd,preprint,superscriptaddress]{revtex4-1}

\usepackage{mathrsfs}
\usepackage{amsfonts}
\usepackage{amsmath,amssymb,bm}
\usepackage[normalem]{ulem}
\usepackage{hyperref}
\hypersetup{colorlinks, linkcolor = [rgb]{0,0.0,0.7}, citecolor = [rgb]{0,0.0,1}, urlcolor = [rgb]{0,0.0,0.7}}
\usepackage{xcolor}
\usepackage{epsfig}
\usepackage{slashed}
\usepackage{hhline,multirow,tabularx}  
\usepackage{url}        
\def\gz{{\gamma Z}}
\def\gw{{\gamma W}}
\def\Re{\textrm{Re}\,}
\def\Im{\textrm{Im}\,}

\begin{document}
\preprint{JLAB-THY-20-3289}

\title{Electroweak axial structure functions and \\ improved extraction of the $V_{ud}$ CKM matrix element}

\author{K.~Shiells}
\thanks{\mbox{Present address: Center for Nuclear Femtography, 1201 New York Ave., NW, Washington DC, 20005, USA}}
\affiliation{University of Manitoba, Winnipeg, MB, Canada R3T 2N2}
\author{P.~G.~Blunden}
\affiliation{University of Manitoba, Winnipeg, MB, Canada R3T 2N2}
\author{W.~Melnitchouk}
\affiliation{Jefferson Lab, Newport News, Virginia 23606, USA}

\begin{abstract}
We present a comprehensive analysis of the $\gw$ interference radiative correction to the neutron \mbox{$\beta$-decay} matrix element.
Within a dispersion relations approach, we compute the axial-vector part of the $\gw$ box amplitude $\Box^{\gw}_{A}$ in terms of the isoscalar part of the $F_3^{\gw}$ interference structure function.
Using the latest available phenomenology for $F_3^{\gw}$ from the nucleon elastic, resonance, deep-inelastic, and Regge regions, we find the real part of the box correction to be \mbox{$\Box^{\gw}_A = 3.90(9) \times 10^{-3}$}. 
This improved correction gives a theoretical estimate of the CKM matrix element $|V_{ud}|^2=0.94805(26)$, which represents a 4$\sigma$ violation of unitarity.
\end{abstract}

\date{\today}
\maketitle

\section{Introduction}

Probing the unitarity of the Cabibbo-Kobayashi-Maskawa (CKM) quark mixing matrix provides a stringent test of the Standard Model of nuclear and particle physics.
The highly dominant CKM matrix element $V_{ud}$ is present in any charged current process involving the coupling between $u$ quarks, $d$ quarks and $W$ bosons.
Any measured charged current cross section with proton and neutron initial or final states will therefore be sensitive to the precise value of $V_{ud}$.  

The empirical relationship between $V_{ud}$ and the measurements performed in superallowed $\beta$-decays is given by~\cite{PDG}
\begin{equation}\label{eq:Vud}
|V_{ud}|^2=\frac{0.97148(20)}{1+\Delta_R^V},
\end{equation}
where $\Delta_R^V$ is a radiative correction term.
Superallowed $\beta$-decays involve transitions between isospin-1 and spin-parity $J^P = 0^+ \to 0^+$ nuclei, and currently provide the most precise extraction of $V_{ud}$, with some 20 accessible superallowed transitions measured over the last 40~years~\cite{Hardy15}.
The measurements of transition energies, half-lives and branching ratios, together with nuclear corrections, determines the numerator of (\ref{eq:Vud}) and its accompanied uncertainty.  The constancy of this quantity amongst the many different superallowed $\beta$ decays, each having their own unique nuclear corrections, has been a testament to the success of these experiments.

Radiative corrections to $\beta$-decays generally fall into two categories: outer corrections, which include energy-dependent terms (such as bremsstrahlung), and inner corrections, which can be computed to high precision and are typically incorporated into effective couplings.
The relevant inner correction $\Delta_R^V$ includes the charged current axial-vector box contribution, denoted by $\Box_A^{\gw}$, involving the exchange of a $W$ boson and a photon between the leptons and hadrons (axial here refers to the coupling of the $W$ to the hadron).
This correction is not protected from the effects of the strong interaction, and has an associated hadronic uncertainty which needs to be accurately estimated.

Several recent attempts have been made to reduce the uncertainties on $V_{ud}$ through more constrained determination of the $\gw$ box diagram~\cite{Marciano06, Czarnecki18, Czarnecki19, Seng18, Seng19}.
Unlike most other one-loop radiative effects in $\beta$-decay, the $\gw$ box contribution depends on details about hadron structure and nonperturbative QCD dynamics.
Consequently, calculations of the $\gw$ box corrections involve modeling the long-distance parts of the $W^+ n \to \gamma p$ amplitude.
In 2006, Marciano and Sirlin~\cite{Marciano06} evaluated the $\gw$ box contribution using a form factor approach, together with nonperturbative hadronic phenomenology, giving a $V_{ud}$ extraction that was consistent with top row CKM unitarity.
More recently, Seng {\it et al.}~\cite{Seng18, Seng19} evaluated the $\gw$ correction using a dispersion relation approach.
This new analysis led to an approximately 15\% increase in the value of the box contribution over that in Ref.~\cite{Marciano06}, leading to an \mbox{$\approx 4\sigma$} deviation from the top row unitarity.
Subsequently, Czarnecki~{\it et al.}~\cite{Czarnecki19} updated the analysis of Ref.~\cite{Marciano06} using improved phenomenological input, resulting in an $\approx 2.5\sigma$ shortfall of unitarity.

In this paper, we perform an independent, comprehensive analysis of the axial-vector part of the $\gw$ box amplitude, $\Box_A^{\gw}$, within a dispersion relation framework, focusing in particular on a systematic assessment of uncertainties arising from nonperturbative QCD physics inputs.
The main uncertainty in the calculation comes from the $F_3^{\gw}(W,Q^2)$ structure function, for which no direct experimental information is available.
(Interference $\gw$ structure functions are not measurable in inclusive DIS, but may be accessible through weak deeply-virtual Compton scattering~\cite{Psaker07}.)
We use model-independent relations between $\gw$ and $\gz$ interference structure functions, as well as input from neutrino and antineutrino inclusive DIS, to provide indirect constraints on $F_3^{\gw}$.

In particular, while the largest contribution to $\Box_A^{\gw}$ comes from the DIS region at large four-momentum transfers $Q^2$ and final state invariant masses $W$, where $F_3^{\gw}$ can be well described in terms of leading-twist parton distribution functions (PDFs), significant strength also comes from the Regge region at high $W$ but low $Q^2$, where the structure function is subject to much greater uncertainty.
Taking all the kinematic regions into account, our analysis gives a total correction $\Box_A^{\gw} = 3.90(9) \times 10^{-3}$, which is higher than recent results~\cite{Czarnecki18, Czarnecki19, Seng18, Seng19}, and leads to a larger discrepancy, $\approx 4\sigma$, with Standard Model unitarity for the top row of the CKM matrix.

This paper is organized as follows.
In Sec.~\ref{sec:gWdisp} we present the theoretical framework for computing the $\gw$ box correction in terms of the interference structure functions using dispersion relations, and give model-independent relations between the $\gw$, $\gz$ and charged current proton and neutron structure functions $F_3^{W^\pm}$.
The nonperturbative inputs into the calculation are described in Sec.~\ref{sec:modeling}, where we discuss the model dependence of the contributions from the nucleon elastic, resonance, Regge and deep-inelastic scattering (DIS) regions.
The numerical results of our calculations for the $\gw$ box correction are presented in Sec.~\ref{sec:results}, and the consequences for the resulting $V_{ud}$ CKM matrix element are discussed in detail.
Finally, in Sec.~\ref{sec:conclusion} we summarize our findings and outline some avenues for future work.
Appendix~\ref{app:F3DIS} contains the details of the derivation of the $F_3^{\gw}$ structure function in the DIS region in terms of leading twist PDFs.

\section{Dispersive approach to the $\gw$ box amplitude}
\label{sec:gWdisp}

\begin{figure}[t]
\centering
\includegraphics[width=0.5\textwidth]{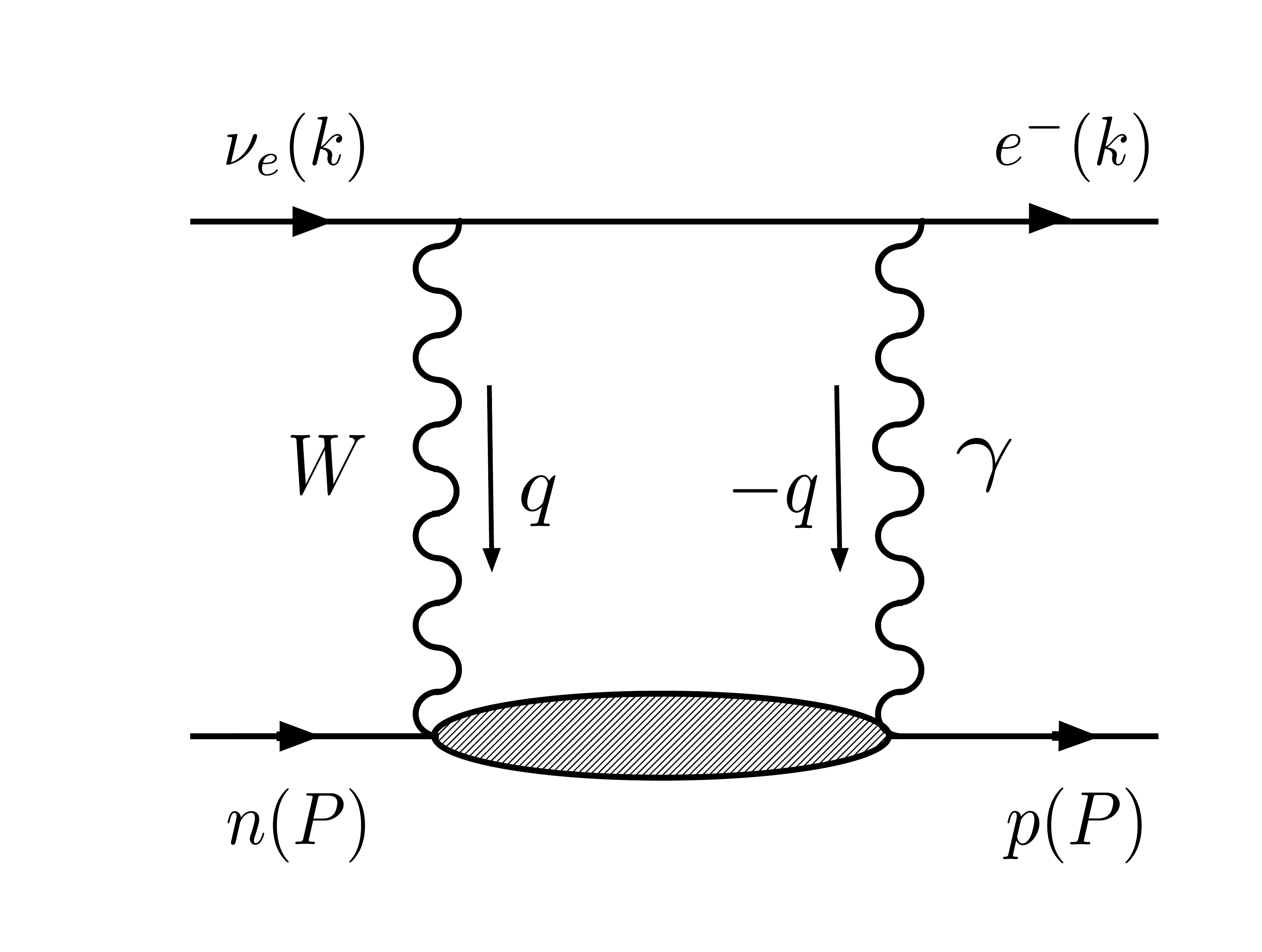}
\vspace*{-0.5cm}
\caption{One-loop electroweak correction to the scattering process
$\nu_e(k) + n (P) \to e^-(k) + p\,(P)$ in the forward limit, involving the absorption of a $W$ boson ($q$) and a virtual photon ($-q$). The corresponding crossed-box diagram is not shown.}
\label{fig:diagram}
\end{figure}

In this section we outline the theoretical framework which we use for the calculation of the one-loop electromagnetic contribution to neutron $\beta$ decay, and describe the methodology of the model-independent dispersive approach employed to compute the correction $\Box_A^{\gw}$.
It is convenient to consider an equivalent forward-angle scattering process, such as $\nu_e\, n\to e^-\, p$, and the associated correction ${\cal M}_\gw$ (shown in Fig.~\ref{fig:diagram}) to the Born amplitude for a pure Fermi transition
\begin{equation}
{\cal M}_W
= V_{ud}\, \left(\frac{G_F}{\sqrt2}\right)\,
  \overline{\psi}_{e}(k) \gamma_\mu (1-\gamma_5) \psi_{\nu_e}(k)\, 2 P^\mu,
\end{equation}
where $G_F$ is the Fermi decay constant.
Following PDG notation~\cite{PDG}, the cross section for scattering of a massless lepton of helicity $\lambda=\pm 1$ can be expressed in terms of products of leptonic and hadronic tensors associated with the coupling of exchanged $\gamma$ and $W$ bosons.
From the optical theorem, the imaginary part of ${\cal M}_\gw$ can be written in terms of the cross section for all possible final hadronic states,
\begin{eqnarray}
2\, \Im {\cal M}_\gw
&=& V_{ud}\, 4\pi\!\int\!\frac{d^3 k'}{(2\pi)^3 2E_{k'}}
\bigg(\frac{4\pi \alpha}{Q^2}\bigg) 
\left(\frac{G_F}{\sqrt{2}}\,\frac{1}{1+Q^2/M_W^2}\right)
L_{\gw}^{\mu\nu} W_{\mu\nu}^\gw\, ,
\end{eqnarray}
where $\alpha$ is the electromagnetic fine structure constant and $M_W$ the $W$ boson mass.
The leptonic tensor is given by
\begin{subequations}
\begin{eqnarray}
L_{\mu\nu}^\gw &=& \left(1 - \lambda\right) L_{\mu\nu}^\gamma,  \\
L_{\mu\nu}^\gamma 
&=& q^2 g_{\mu\nu} + 4 k_\mu k_\nu - 2 k_\mu q_\nu - 2 k_\nu q_\mu
 - i\lambda \epsilon_{\mu\nu}(k q),
\end{eqnarray}
\end{subequations}
where $q=k-k'$ is the virtual four-momentum transfer, with $Q^2 \equiv -q^2$, and we use the notation $\epsilon_{\mu\nu}(k q) \equiv \epsilon_{\mu\nu\alpha\beta}\, k^\alpha q^\beta$, with $\epsilon_{0123}=-1$.

The hadronic tensor is given by
\begin{eqnarray}
W_{\mu\nu}^\gw
&=& \frac{1}{4\pi} \int d^4 x\,e^{i q\cdot x}\ \langle p \left| T \left[ J_\mu^W(x) J_\nu^\gamma(0)\right] \right| n\rangle,
\end{eqnarray}
where the electromagnetic and 
charge-raising weak
currents can be written (for 3 flavors) in terms of quark currents as
\begin{subequations}
\begin{eqnarray}
J_\mu^\gamma\,
&=& e_u\, \bar{u}\gamma_\mu u + e_d\, \bar{d} \gamma_\mu d + e_s\, \bar{s} \gamma_\mu s, \\
J_\mu^W &=& \bar{u}\gamma_\mu (1-\gamma_5) d.
\end{eqnarray}
\end{subequations}
The tensor $W_{\mu\nu}^\gw$ is usually decomposed in terms of the interference electroweak structure functions $F_i^\gw$,
\begin{equation}
W_{\mu\nu}^\gw = - g_{\mu\nu} F_1^\gw
  + \frac{P_\mu P_\nu}{P\cdot q} F_2^\gw
  - i \frac{\epsilon_{\mu\nu}(P q)}{2 P\cdot q} F_3^\gw.
\end{equation}
Our focus here is on the $F_3^\gw$ contribution, involving the axial-vector hadronic coupling of the $W$ boson. 
Making a change of variables
\begin{equation}
\frac{d^3 k'}{(2\pi)^3 2 E_{k'}}\
\to\ \frac{1}{32 \pi^2\, k\!\cdot\!P}\, dW^2\, dQ^2\, ,
\end{equation}
setting $\lambda=-1$, and evaluating $k\!\cdot\!P = M E$ in the rest frame of the neutron, we find
\begin{eqnarray}
\Im {\cal M}_\gw(E)
&=& \frac{\alpha G_F}{\sqrt{2} M E}
\int\limits_{M^2}^s \! dW^2 \!\!\int\limits_0^{Q_{\rm max}^2}\!\! dQ^2\,
\frac{F_3^\gw(W^2,Q^2)}{1+Q^2/M_W^2} \, 
\left(\frac{2 M E}{W^2-M^2+Q^2} - \frac{1}{2}\right) ,
\end{eqnarray}
where $s = (k+P)^2 = M^2+2 M E$, and $Q^2_{\rm max} = 2ME (1-W^2/s)$.
The structure function $F_3^\gw(W^2,Q^2)$ is a function of the invariant mass squared, $W^2$, of the exchanged boson and hadron and of the exchanged boson virtuality, $Q^2$.

Defining the correction $\Box^\gw_A$ to the Born amplitude via
\begin{equation}
{\cal M}_W + {\cal M}_\gw\ \to\ {\cal M}_W\, \big(1+\Box^\gw_A\big),
\end{equation}
the correction to $\Delta_R^V$ from the $\gw$ box is therefore given by $2\, \Box^\gw_A$.
After some elementary trace algebra, we find for the imaginary part of $\Box^\gw_A$:
\begin{eqnarray}
\Im \Box^\gw_{A}(E) 
&=& \frac{\alpha}{32(M E)^2}
\int\limits_{M^2}^s \! dW^2 \!\!\int\limits_0^{Q_{\rm max}^2}\!\!dQ^2\,
\frac{F_3^\gw(W^2,Q^2)}{1+Q^2/M_W^2}
\left(\frac{2 M E}{W^2-M^2+Q^2} - \frac{1}{2}\right).
\end{eqnarray}
It has been known for some time~\cite{Sirlin67, Sirlin78, Towner92} that the isovector part of the electromagnetic current does not contribute to the $\gw$ box in a direct loop integral approach at zero energy, such as for $\beta$-decay.
In the dispersive formulation, Seng~{\it et al.}~\cite{Seng18, Seng19} showed that the isoscalar and isovector electromagnetic currents for the $\gw$ box scattering amplitude have opposite crossing symmetries under $E \to -E$.
In terms of $\Box_A^\gw(E)$, this results in the dispersion relations
\begin{subequations}
\label{eq:disp}
\begin{eqnarray}
\Re \Box_A^{\gw (0)}(E)
&=& \frac{2}{\pi}\int_0^\infty dE'
\frac{E'}{E'^2-E^2}\ 
\Im \Box_A^{\gw (0)}(E'),\\
\Re \Box_A^{\gw (1)}(E)
&=& \frac{2 E}{\pi}\int_0^\infty dE'
\frac{1}{E'^2-E^2}\ 
\Im \Box_A^{\gw (1)}(E'),
\end{eqnarray}
\end{subequations}%
where the superscripts $(0)$ and $(1)$ refer to the isoscalar and isovector electromagnetic current contributions to $F_3^\gw$, which are denoted as $F_3^{(0)}$ and $F_3^{(1)}$, respectively, in keeping with the notation of Seng~{\it et al.}~\cite{Seng19}.
From Eqs.~(\ref{eq:disp}) it is clear that the real part of the isovector contribution vanishes at zero energy, $\Re \Box_A^{\gw (1)}(E=0)=0$.
To avoid unnecessary clutter in our notation, we will henceforth denote the real part of the isoscalar contribution,
$\Re \Box_A^{\gw (0)}(E=0)$, as simply $\Box_A^\gw$.

Following the approach of earlier work on the $\gz$ interference correction $\Box_A^\gz$~\cite{Blunden11, Blunden12}, the triple integral for $\Box_A^\gw$ can be simplified by changing the order of integration so that the energy integral can be performed analytically. 
A further change of variable from $W^2$ to the Bjorken scaling variable $x=Q^2/(W^2-M^2+Q^2)$ gives the compact expression
\begin{eqnarray}
\Box^{\gw}_A
&=& \frac{\alpha}{2 \pi}
    \int_0^\infty dQ^2\,
    \frac{1}{Q^2(1+Q^2/M_W^2)}
    \int_0^1\!\! dx\ F_3^{(0)}(x,Q^2)\, \frac{1+2 r}{(1+r)^2},
\label{eq:boxRe}
\end{eqnarray}
with $r=\sqrt{1 + 4 M^2 x^2/Q^2}$.
Written in this way, the integrand can be expanded in a series of Nachtmann moments of the structure function $F_3^{(0)}$~\cite{Blunden11,Seng18}. At high $Q^2$, the leading order lowest moment is related to the Gross-Llewellyn-Smith (GLS) sum rule, and is independent of hadronic structure.

Before we continue with modeling $F_3^{(0)}(x,Q^2)$ in different kinematic regions, it is worthwhile to note the relationship with other axial-vector interference structure functions. 
Seng~{\it et al.}~\cite{Seng18, Seng19} demonstrated that the $\Box_A^\gw$ box correction is related through isospin symmetry to the axial $\Box_A^\gz$ box corrections studied in earlier work on parity-violating electron-proton scattering~\cite{Blunden11, Blunden12, Rislow13}.
In our notation, they give the relations
\begin{subequations}
\label{eq:F3relations}
\begin{eqnarray}
F_3^{(0)}
&=& F_{3p}^{\gz} - F_{3n}^{\gz},
\label{F30relation}    \\
F_3^W &\approx& F_{3p}^{\gz}+F_{3n}^{\gz},
\label{F3Wrelation}
\end{eqnarray}
\end{subequations}
where we define $F_3^W \equiv \frac{1}{2} \big( F_3^{W^+} + F_3^{W^-} \big)$ in terms of the axial $\nu p$ and $\bar\nu p$ charged current structure functions $F_3^{W^+}$ and $F_3^{W^-}$, respectively~\cite{PDG}.
To be consistent with PDG conventions, our definition of $F_3^{(0)}$ is four times larger than that used in Refs.~\cite{Seng18, Seng19}.
Equation~(\ref{F30relation}) is a consequence of isospin symmetry, while Eq.~(\ref{F3Wrelation}) is only an approximate equality due to the different contributions of strange and heavier quarks to $F_3^{W^\pm}$ and $F_{3}^{\gz}$.

\section{$\gw$ interference structure function}
\label{sec:modeling}

The most uncertain part of the calculation of the $\gw$ box correction in Eq.~(\ref{eq:boxRe}) is the interference structure function $F_3^{(0)}$. The precision to which this function can be estimated will ultimately determine the precision of the $\gw$ correction to the $V_{ud}$ matrix element.
Since the empirical and theoretical information on $F_3^{(0)}$ depends somewhat on kinematics, it is useful to consider the structure function in four specific regions of $W^2$ and $Q^2$:
\begin{enumerate}
\itemsep0em
\item[(i)] elastic (``el''), 
for $W^2 = M^2$;
\item[(ii)] deep-inelastic scattering (``DIS''),
for $W^2 \geq W_0^2$ and $Q^2 \geq Q_0^2$;
\item[(iii)] Regge (``Reg''), 
for $W^2 \geq W_0^2$ and $Q^2 \leq Q_0^2$;
\item[(iv)] resonances (``res'') + background (``bgd'')
for $W_\pi^2 \le W^2 \leq W_0^2$, where $W_\pi = M + m_\pi$.
\end{enumerate}

\begin{figure}[t]
\includegraphics[width=0.75\textwidth]{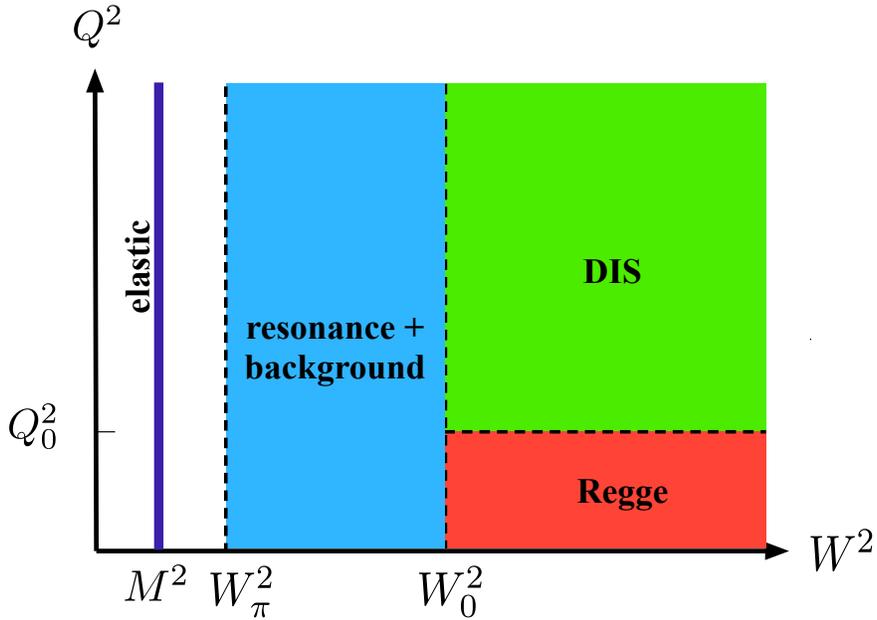}
\vspace*{-0.8cm}
\caption{Kinematical regions of $Q^2$ and $W^2$ into which contributions to the $F_3^{(0)}$ structure function are separated: nucleon elastic ($W^2=M^2$), resonance + background ($W_\pi^2 \leq W^2 \leq W_0^2$), Regge ($W^2 >  W_0^2$; $Q^2 < Q_0^2$), and DIS ($W^2 > W_0^2$; $Q^2 > Q_0^2$).}
\label{fig:regions}
\end{figure}

These regions are illustrated in Fig.~\ref{fig:regions}.
In our analysis we 
use as our nominal boundary separating the DIS and Regge regions the value $Q_0^2=2$~GeV$^2$, but vary this between $Q_0^2=1$ and 2~GeV$^2$ to test the stability of the results.
The boundary between the low-$W$ nucleon resonance and high-$W$ DIS and Regge regions is set to the traditional value of $W_0^2=4$~GeV$^2$.
In this section we discuss in detail the existing constraints on $F_3^{(0)}$ in the various regions, and estimate their uncertainties.

\subsection{Elastic contribution}
\label{ssec:elastic}

The contributions to the $F_3^{(0)}$ structure function from region {(i)} can be written in terms of the elastic electroweak form factors as~\cite{Blunden11, Blunden12, Seng18, Seng19}
\begin{equation}
F_{3\, (\text{el})}^{(0)}(Q^2)
= -\big[ G_M^p(Q^2) + G_M^n(Q^2) \big]\,G_A(Q^2)\,Q^2 \delta(W^2-M^2).
\end{equation}
For use in Eq.~(\ref{eq:boxRe}), we note that $Q^2\,\delta(W^2-M^2)=x\,\delta(1-x)$.
For the elastic magnetic form factors of the proton and neutron we use the recent parametrization from Ref.~\cite{Arrington18}, which accounts for two-photon exchange effects in its extraction.

The largest contributor to the uncertainty in the elastic contribution is the $Q^2$ dependence of the axial-vector form factor, $G_A(Q^2)$.
We consider the two-component parametrization for $G_A$ given by Megias {\it et al.}~\cite{Megias20},
\begin{eqnarray}
\label{eq:GA}
G_A(Q^2)
&=& -\frac{g_A}{(1+c_1 Q^2)^2}
    \bigg( 1 - c_2 + c_2 \frac{m_A^2}{m_A^2+Q^2} \bigg),
\end{eqnarray}
where $g_A=1.2756(13)$ is the nucleon axial coupling~\cite{PDG}, $m_A=1.23(4)$~GeV is taken to be the mass of the axial $a_1(1260)$ meson, and $c_1$ and $c_2$ are fitting parameters. Setting $c_2=0$ and $c_1=1/m_A^2$ gives the commonly used dipole form factor parametrization, with $m_A$ now taken to be a free parameter.

Historically, the world average of simple dipole fits to $G_A(Q^2)$ from both neutrino scattering and electroproduction experiments has been established by Bernard {\it et al.}~\cite{Bernard02}.
There is still ongoing debate on the value of the axial mass parameter $m_A$, and whether neutrino scattering or electroproduction is a more accurate method for its extraction.
An attempt was made by Bhattacharya {\it et al.}~\cite{Bhattacharya11} to analyze the axial form factor in a model-independent approach, with comparisons to previous work, and this was ultimately used in the analysis of Seng {\it et al.}~\cite{Seng18, Seng19}.
The form of Eq.~(\ref{eq:GA}) is motivated by the attempt to incorporate pion loop corrections to the axial form factor, and can be applied to either neutrino scattering or pion electroproduction data with its fitting parameters $c_1$ and $c_2$.  

\begin{figure}[t]
\includegraphics[width=0.75\textwidth]{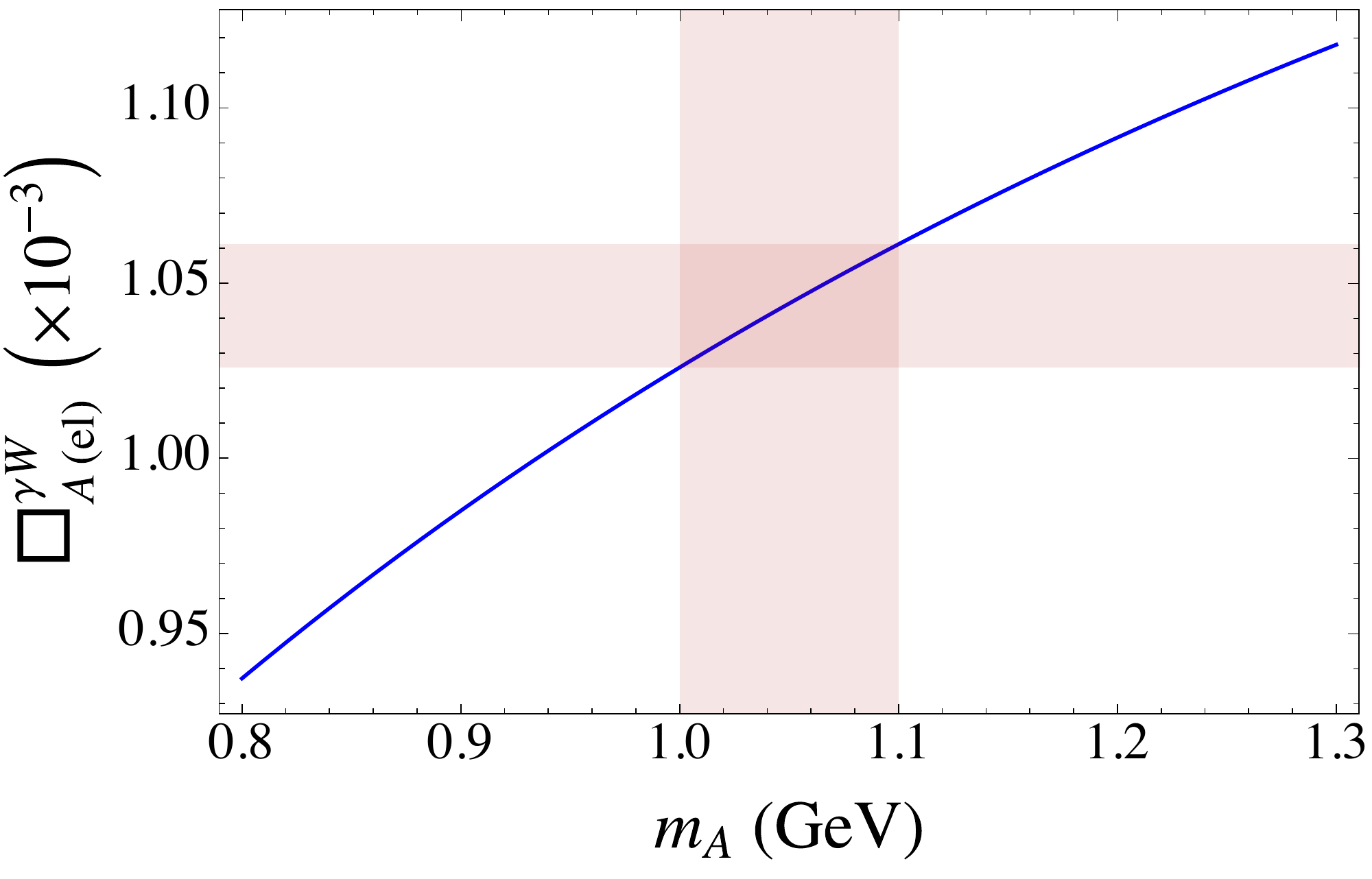}
\caption{Dependence of the elastic intermediate state contribution $\Box_{A\, (\rm el)}^{\gw}$ to the $\gw$ box diagram on the dipole axial mass parameter $m_A$ (solid blue line), including the 1$\sigma$ uncertainty band for $m_A=1.05(5)$~GeV (pink bands).}
\label{fig:boxel}
\end{figure}

Taking into consideration all of the possible $(c_1,c_2)$ values given in Ref.~\cite{Megias20}, the elastic contribution is found to be
$\Box^{\gw}_{A\, \rm (el)} = 1.05(3) \times 10^{-3}$,
where the uncertainty is two parts systematic and one part statistical.
The sensitivity of $\Box^{\gw}_{A\, \rm (el)}$ to the the axial mass parameter $m_A$ for a dipole form factor is illustrated in Fig.~\ref{fig:boxel} over the range of possible values in use.
Using the form of $G_A(Q^2)$ from Ref.~\cite{Bhattacharya11} leads to a nearly equivalent central value, but with twice the uncertainty.

One can also conservatively take the average of the two $m_A$ world averages in Ref.~\cite{Bernard02} by setting $m_A=1.05(5)$~GeV.
The result of doing so again reveals a nearly equivalent $\Box^{\gw}_{A\, \rm (el)}$ central value as that obtained using the $G_A(Q^2)$ of Megias {\it et al.}~\cite{Megias20} and Bhattacharya {\it et~al.}~\cite{Bhattacharya11}.
Since all three approaches~\cite{Bernard02, Bhattacharya11, Megias20} give results that are 
consistency, we opt to use the central value from Megias {\it et al.}~\cite{Megias20}, and increase the uncertainty slightly to reflect the limitations of the other approaches.
We therefore take
    $\Box^{\gw}_{A\, \rm (el)} = 1.05(4) \times 10^{-3}$
as the main result for the elastic intermediate state contribution in our analysis.

\subsection{DIS contribution}
\label{ssec:DIS}

The contribution to $\Box^{\gw}_A$ from the DIS region {(ii)} can be writtem as
\begin{eqnarray}
\Box^{\gw}_{A\,  (\rm DIS)}
&=& \frac{1}{2\pi}
    \int_{Q_0^2}^\infty dQ^2
    \frac{\alpha(Q^2)}{Q^2(1+Q^2/M_W^2)}
    \int_0^{x_0}\!\! dx\ F_{3\, (\rm DIS)}^{(0)}(x,Q^2)\,
    \frac{1+2 r}{(1+r)^2}\, ,
\label{eq:F3DIS}					
\end{eqnarray}
where $x_0 = Q^2/(W_0^2-M^2+Q^2)$ is the upper limit on the $x$ integration corresponding to the minimum value of $W^2$ given by $W_0^2$ (see Fig.~\ref{fig:regions}).
Setting $W_0^2 = 4$~GeV$^2$ excludes contributions from the nucleon resonance region, which cannot be described in terms of structure functions evaluated from PDFs.
The lower $Q^2$ boundary of the DIS region, denoted by $Q_0^2$, ensures that the integral (\ref{eq:F3DIS}) is dominated by the leading twist part of the $\gw$ structure function, with negligible ${\cal O}(1/Q^2)$ power corrections from higher twist contributions.
In our numerical analysis we explore the stability of the results with respect to variations of the boundary between $Q_0^2 = 1$ and 2~GeV$^2$.

To account for the large variation of $\alpha(Q^2)$ in DIS kinematics, we follow the practice introduced in the corresponding $\Box_A^\gz$ calculations~\cite{Blunden11, Blunden12} and include the running of $\alpha(Q^2)$ under the $Q^2$ integral in Eq.~(\ref{eq:F3DIS}).
To do this we use the parametrization of Jegerlehner~\cite{Jegerlehner17}, which partially accounts for two-loop effects in the photon propagator, and results in a 4\% enhancement over using a fixed $\alpha(0)$.
Equivalently, this corresponds to using a constant value of $\alpha$ set at a scale $Q^2 \approx 12.8$~GeV$^2$, which is close to the weighted average of $Q^2$ in the integrand of (\ref{eq:F3DIS}).

\begin{figure}[t]
\includegraphics[width=0.75\textwidth]{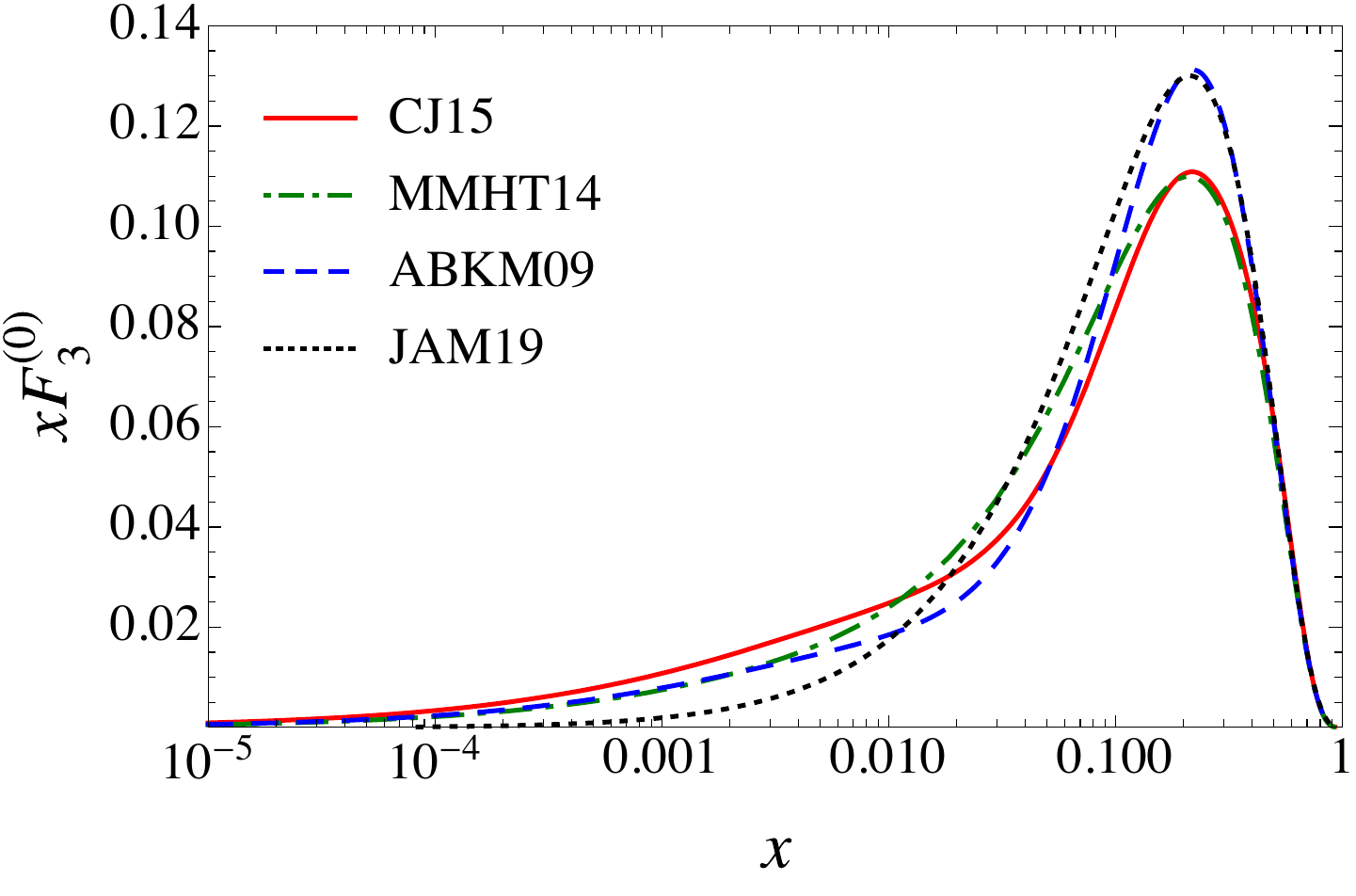}
\vspace*{-0.5cm}
\caption{Comparison of the $x F_{3\, (\rm DIS)}^{(0)}$ structure function versus $x$ for several PDF parametrizations at fixed $Q^2=9$~GeV$^2$, from the CJ15~\cite{CJ15} (solid red line), MMHT14~\cite{MMHT14} (dot-dashed green line), ABKM09~\cite{ABKM09} (dashed blue line) and JAM19~\cite{JAM19} (dotted black line) global QCD analyses.}
\label{fig:F3DIS}
\end{figure}

In the DIS region, where perturbative QCD is applicable, the $F_{3\, \rm(DIS)}^{(0)}$ structure function can be written in factorized form as a convolution of valence $u_v \equiv u-\bar u$ and $d_v \equiv d-\bar d$ quark PDFs and the nonsinglet Wilson coefficient function (see Appendix~\ref{app:F3DIS} for details of the derivation),
\begin{equation}\label{eq:F30DIS}
F^{(0)}_{3\, (\rm DIS)}(x,Q^2)
= \frac13 \int_x^1 \frac{dz}{z}\, C_3(x/z,Q^2) \big(u_v-d_v\big)(z,Q^2),
\end{equation}
where we have set $e_u+e_d=1/3$.
The coefficient function describes the hard scattering of the virtual photon from the parton, and up to ${\cal O}(\alpha_s^2)$ corrections is given by
\begin{eqnarray}
C_3(z,Q^2)
&=& \delta(1-z)
 +  \frac{\alpha_s(Q^2)}{4\pi}
\bigg[ 2(1+z)\, \ln\frac{z}{1-z}
     - \frac{4 \ln z}{1-z}
     + 2(z+2)
\nonumber\\
& &
 - \Big( \frac{2\pi^2}{3} + 9 \Big) \delta(1-z)
 + \bigg( \frac{4 \ln(1-z)}{1-z} \bigg)_+
 - \frac{3}{(1-z)_+} 
\bigg]\
+\ {\cal O}(\alpha_s^2),
\label{eq:C3}
\end{eqnarray}
where the ``+'' terms are understood as distributions which are made finite in the limit as $x \to 1$ inside any integral over $x$.
The $x$ dependence of $x F_{3\, (\rm DIS)}^{(0)}$ is illustrated in Fig.~\ref{fig:F3DIS} for different sets of PDFs obtained by several groups from global QCD analyses~\cite{CJ15, MMHT14, ABKM09, JAM19}.
Since the valence $u$ and $d$ quark distributions are fairly well constrained experimentally, the differences between the various parametrizations are relatively small, with $x F_{3\, (\rm DIS)}^{(0)}$ peaking around $x \approx 0.2$, and dropping rapidly as $x \to 0$.

In the high-$Q^2$ limit, when both $x_0 \to 1$ and $r \to 1$, the integral over $x$ in Eq.~(\ref{eq:F3DIS}) takes the simple form
\begin{equation}
\label{eq:F30int}
\int_0^1 dx\, F^{(0)}_{3\, (\rm DIS)}(x,Q^2)
= \frac13 \bigg( 1 - \frac{\alpha_s(Q^2)}{\pi} + {\cal O}\big(\alpha_s^2\big) \bigg),
\end{equation}
so that the area under each of the curves in Fig.~\ref{fig:F3DIS} is fixed at a given $Q^2$.
Higher order corrections to Eq.~(\ref{eq:F30int}) have been computed to ${\cal O}(\alpha_s^4)$~\cite{Larin91, Baikov10} and used in Ref.~\cite{Czarnecki19} in their calculation of $\Box^{\gw}_A$.
On the other hand, the approximation in Eq.~(\ref{eq:F30int}) assumes negligible contributions from the region $x > x_0$ in Eq.~(\ref{eq:F3DIS}), which can introduce errors into the calculation.
In our analysis we therefore compute the integrals in (\ref{eq:F3DIS}) exactly at finite $Q^2$, using the explicit form of the ${\cal O}(\alpha_s)$ correction in Eq.~(\ref{eq:C3}), and quantify the omitted large-$x$ contribution explicitly.

We also note that Ref.~\cite{Seng19} gives the $\Box_A^{\gw}$ correction as being proportional to the lowest Nachtmann moment of $u-\bar{d}$.
This disagrees with the structure function relationship in Eq.~(\ref{F30relation}), which is quite general.
In fact, such a combination of PDFs is equal to $u_v + (\bar{u}-\bar{d})$, from which the second term is non-negligible.
To make matters worse, the distribution of $\bar{u}-\bar{d}$ is poorly constrained within globally determined PDFs, and its large uncertainty would overwhelm the sought precision of the $V_{ud}$ extraction (see Appendix~\ref{app:F3DIS}).

\begin{figure}[t]
\centering
\includegraphics[width=0.75\textwidth]{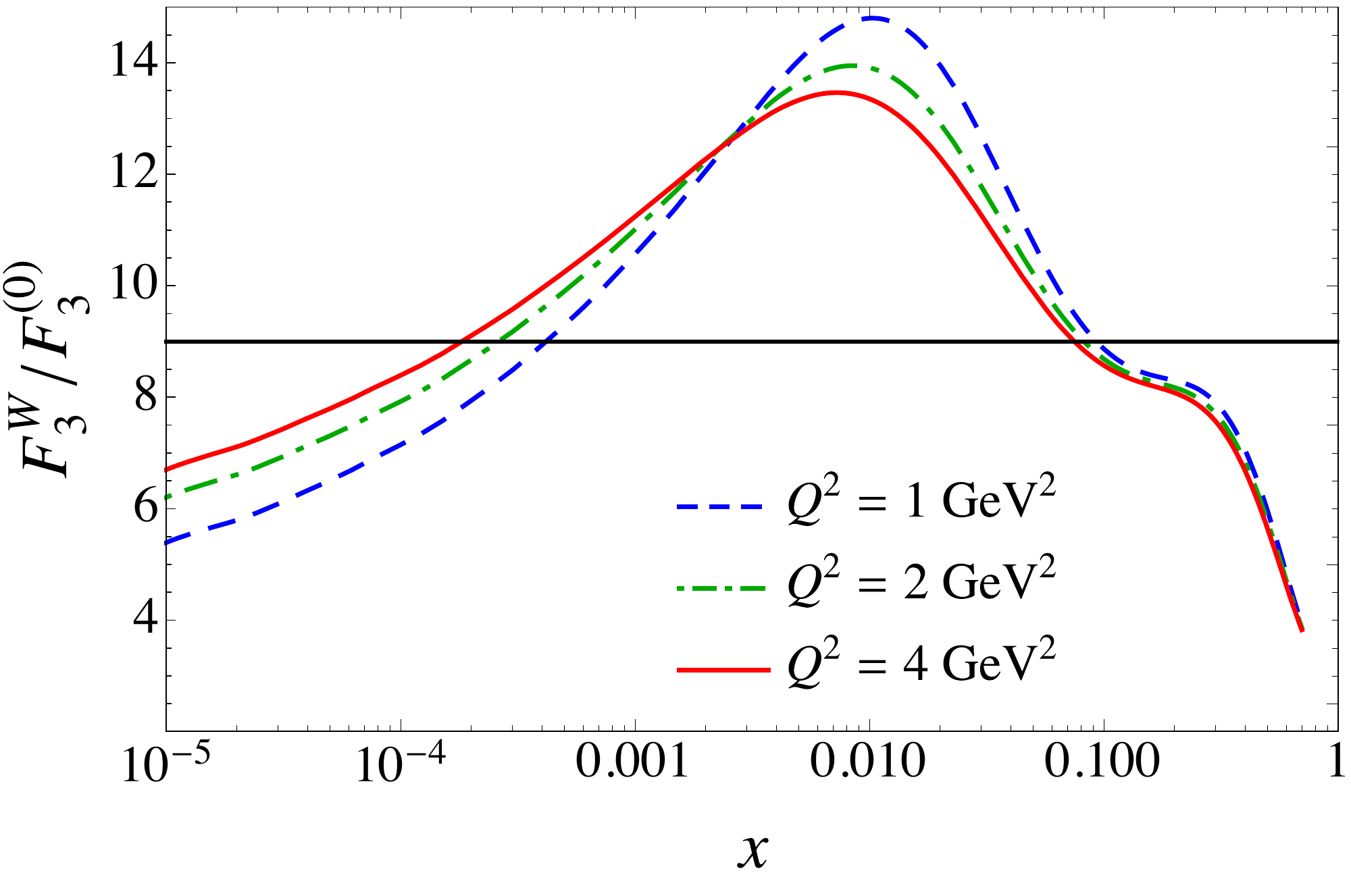}
\vspace*{-0.5cm}
\caption{Comparison of the $F^W_{3\, (\rm DIS)}/F^{(0)}_{3\, (\rm DIS)}$ structure function ratio versus $x$ for fixed values of $Q^2=1$ (dashed blue line), 2 (dot-dashed green line) and 4~GeV$^2$ (solid red line) using the PDF parametrizations from Ref.~\cite{MMHT14}, and compared with the constant ratio of 9 (horizontal black line) motivated by the ratio of integrated functions, Eqs.~(\ref{eq:F30int}) and (\ref{eq:F3Wint}).}
\label{fig:f3wf30}
\end{figure}

Since we do not have direct empirical information on $\gw$ interference structure functions, it was suggested by Seng~{\it et al.}~\cite{Seng18, Seng19} to use Eqs.~(\ref{eq:F3relations}) to relate $F_3^{(0)}$ to the $F_3^{W^\pm}$ structure functions measured in inclusive neutrino scattering, for which data do exist, albeit not very precise and only on nuclear targets.
In particular, Ref.~\cite{Seng19} considered the integrated value of $F^{(0)}_{3\, (\rm DIS)}$ as in Eq.~(\ref{eq:F30int}) and related this to the integrated value of the $F^W_{3\, (\rm DIS)}$ structure function, which is given by the GLS sum rule as the lowest moment of the valence $u_v+d_v$ distributions in the proton,
\begin{equation}
\label{eq:F3Wint}
\int_0^1 dx\, F^W_{3\, (\rm DIS)}(x,Q^2)
= 3\, \bigg( 1 - \frac{\alpha_s(Q^2)}{\pi} + {\cal O}\big(\alpha_s^2\big) \bigg).
\end{equation}
Furthermore, they assumed that the ratio of 9 between Eqs.~(\ref{eq:F30int}) and (\ref{eq:F3Wint}) also holds as a function of $x$.
In Fig.~\ref{fig:f3wf30} we explore this assumption by examining the $x$ dependence of the ratio $F^W_{3\, (\rm DIS)}/F^{(0)}_{3\, (\rm DIS)}$, computed from the PDF parametrizations of Ref.~\cite{MMHT14} for several fixed values of $Q^2$ from 1 to 4~GeV$^2$.
Compared with the constant ratio of 9, which is also shown for reference, the calculated ratio exceeds this by up to $\approx 50\%$ at intermediate values of $x \sim 0.1$, but underestimates it at low ($x \lesssim 10^{-4}$) and high ($x \gtrsim 0.1$) $x$ values, such that the ratio of the integrated strengths of $F^W_{3\, (\rm DIS)}$ and $F^{(0)}_{3\, (\rm DIS)}$ averages to around 9.
In our numerical analysis below, we consider both scenarios, in which $F^{(0)}_{3\, (\rm DIS)}$ is computed entirely from leading twist PDFs, and also where it is related to the charged current structure functions by the constant overall factor of 9.

Putting these effects together, we find that 
for $Q_0^2=2$~GeV$^2$ 
the contribution from the DIS region (ii) to the $\gw$ box correction is $\Box^{\gw}_{A\, \rm (DIS)} = 2.29(3)\times 10^{-3}$, which is $\sim 2$ times larger than the elastic contribution in Sec.~\ref{ssec:elastic}.
Setting the boundary at $Q_0^2=1$~GeV$^2$ would result in an $\approx 6\%$ increase in the correction, which, however, would be compensated somewhat by a larger contribution from the Regge region to be discussed in Sec.~\ref{ssec:Regge} below. 
The total uncertainty on this estimate is a quadrature sum of a 1\% uncertainty from omitted higher-order perturbative QCD contributions, and a 0.5\% uncertainty associated with the numerical value of the strong coupling, $\alpha_s$.
The main difference between our determination of $\Box^{\gw}_{A\, \rm (DIS)}$ and that of Refs.~\cite{Seng18, Seng19} is in the latter's assumption of the validity of perturbative QCD over all $x$ values up to $x=1$, and subsequent use of perturbative QCD sum rules which have corrections computed to higher orders.
In practice, however, the difference between the two approaches is found to be fairly small for the final value of $\Box^{\gw}_{A\, \rm (DIS)}$.

\subsection{Regge contribution}
\label{ssec:Regge}

While some empirical guidance is available for the DIS and elastic contributions to the $F_3^{(0)}$ structure function, there are even fewer direct constraints in the Regge region (iii) at high $W^2$ ($W^2 \geq W_0^2$) and low $Q^2$ ($Q^2 \leq Q_0^2$), corresponding to low values of $x$.
Some data do exist on the charged current $F_3^W$ structure function from bubble-chamber neutrino and antineutrino scattering experiments at CERN~\cite{Bolognese83, Bosetti82} in the range $0.1 \lesssim Q^2 \lesssim 100$~GeV$^2$.

In Ref.\cite{Seng19}, Seng {\it et al.} use a Regge model which includes the exchange of two vector mesons, identified with $\rho$ and $a_1$ mesons.
Fitting to the Nachtmann moments of the $F_3^W$ structure function data~\cite{Bosetti82}, they obtain reasonably good descriptions of the $x$-integrated data, albeit within large uncertainties.
In our present approach, we use the full information from the same experiment, on both the $x$ and $Q^2$ dependence of the data, in order to better constrain the $F_3^{(0)}$ structure function.
We use a hybrid Regge model for the nonsinglet structure function which builds in the known $x$ dependence of the valence quark distributions at low $x$ (high $W$), together with an interpolation mapping on to the expected behavior as $Q^2 \to 0$ and $Q^2 \to Q_0^2$.

Such a behavior is realized in the Regge model of Capella {\it et al.}~\cite{Capella94}, in which the charged current neutrino structure function is parametrized by the form
\begin{eqnarray}
F_{3\, \rm (Reg)}^W(x,Q^2)
&=& A^{p+n}\, x^{-\alpha_R}(1-x)^c
    \bigg( \frac{Q^2}{Q^2+\Lambda_R^2} \bigg)^{\alpha_R}.
\label{eq:F3WReg}
\end{eqnarray}
For the small-$x$ exponent $\alpha_R$ we use the value $\alpha_R = 0.477$ from Ref.~\cite{Capella94} obtained from Regge model descriptions of high-energy photon-hadron and hadron-hadron scattering data.
Fitting the data from Ref.~\cite{Bosetti82} to the form (\ref{eq:F3WReg}), with the constraint that $F_{3\, \rm (Reg)}^W(x,Q_0^2) = F_{3\, \rm (DIS)}^W(x,Q_0^2)$ along the boundary $Q_0^2=2$~GeV$^2$, we find good agreement with the data for values of the large-$x$ exponent $c = 0.63(1)$, the low-$Q^2$ interpolation mass parameter $\Lambda_R = 0.50(7)$, and the normalization $A^{p+n} = 2.22(4)$.
The constraint from the matching with the DIS region at $Q^2=Q_0^2$ is significant, partly due to the large relative weight of the DIS data arising from the small uncertainties on the valence quark PDFs from global QCD analyses.
In practice, we find that the $\gw$ box correction is rather insensitive to the value of $c$, as may be expected for low-$x$ dominance of this contribution, but is moderately sensitive to both $A^{p+n}$ and $\Lambda_R$.

Extending the model (\ref{eq:F3WReg}) to the isoscalar electromagnetic current, we parametrize the Regge contribution to the $F_3^{(0)}$ structure function by an analogous form,
\begin{eqnarray}
F_{3\, \rm (Reg)}^{(0)}(x,Q^2)
= A^{p-n}\, x^{-\alpha_R} (1-x)^c\,
  \bigg( \frac{Q^2}{Q^2+\Lambda_R^2} \bigg)^{\alpha_R},
\label{eq:F30Reg}
\end{eqnarray}
assuming that the exponent $c$ is the same for protons and neutrons, and the $Q^2 \to 0$ behavior of $F_{3\, \rm (Reg)}^{(0)}$ is the same as $F_{3\, \rm (Reg)}^W$.
For the normalization $A^{p-n}$ in Eq.~(\ref{eq:F30Reg}) we use the ratios depicted in Fig.~\ref{fig:f3wf30} to obtain $F_{3\, \rm (Reg)}^{(0)}$ for $Q^2 \geq Q_0^2$ and $W^2 \geq W_0^2$ with either the constant proportionality
\begin{equation}
A^{p-n} = \frac19\, A^{p+n},
\label{eq:Apn_const}
\end{equation}
or the $x$-dependent ratio at the $Q^2=Q_0^2$ boundary,
\begin{equation}
A^{p-n} = A^{p+n}\ \frac{F_{3\, \rm (Reg)}^{(0)}}{F_{3\, \rm (Reg)}^W}\Bigg|_{Q^2=Q_0^2}\, .
\label{eq:Apn_dyn}
\end{equation}
For the central value of our calculation we use the constant proportionality in Eq.~(\ref{eq:Apn_const}), and estimate the systematic uncertainty on this by taking the difference with the result using the $x$-dependent ratio in Eq.~(\ref{eq:Apn_dyn}).

\begin{figure}[t]
\centering
\includegraphics[width=0.75\textwidth]{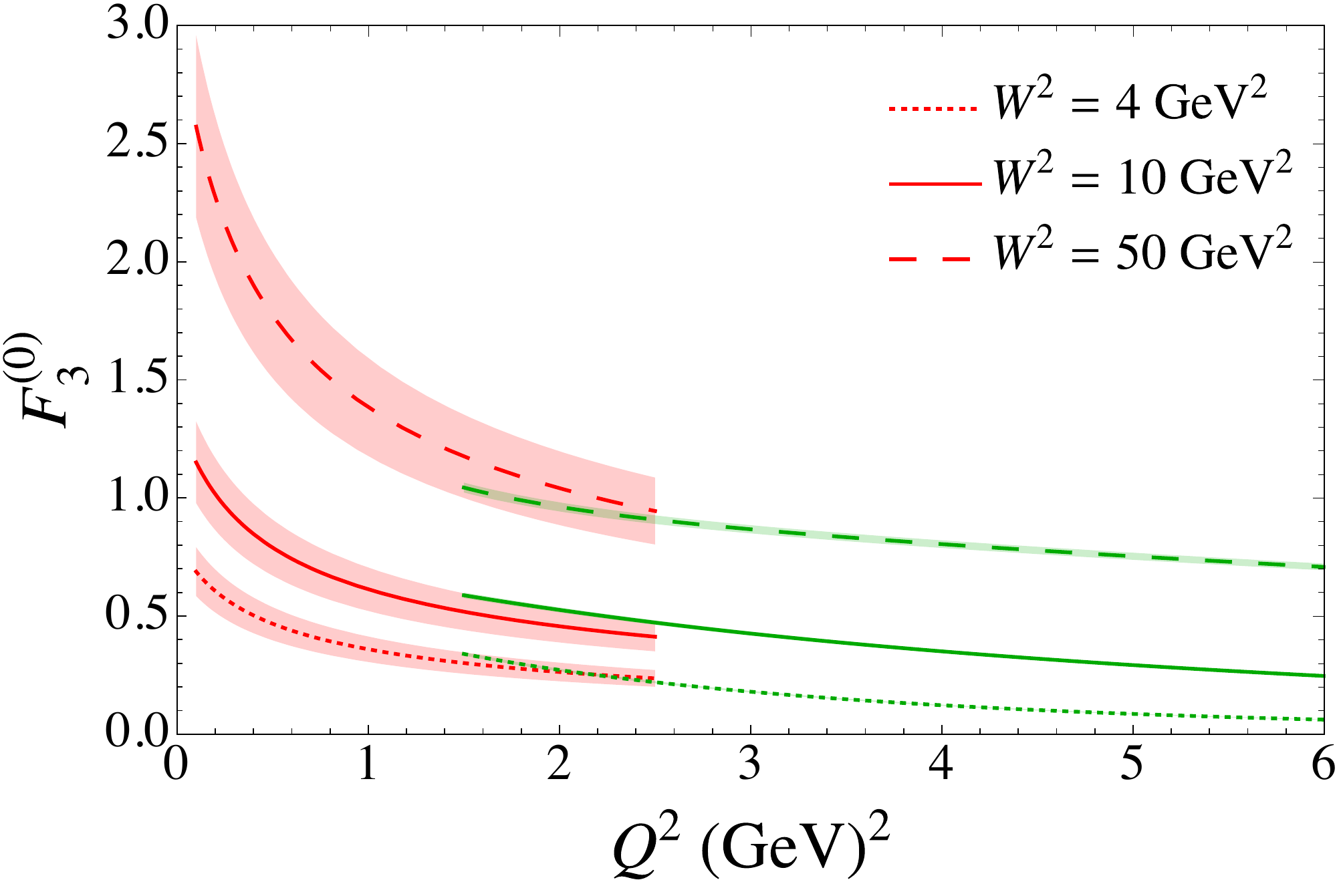}
\caption{Matching of the $F_{3}^{(0)}$ structure function between the Regge region (red lines and bands) and DIS region (green lines and bands) versus $Q^2$ for fixed $W^2=4$~GeV$^2$ (dotted), 10~GeV$^2$ (solid) and 50~GeV$^2$ (dashed).}
\label{fig:matching_Reg_DIS}
\end{figure}

The resulting matching between the structure functions in the Regge region, $F_{3\, \rm (Reg)}^{(0)}$, and the DIS region, $F_{3\, \rm (DIS)}^{(0)}$, is illustrated in Fig.~\ref{fig:matching_Reg_DIS} as a function of $Q^2$, for several fixed values of $W^2$ from $W^2=4$~GeV$^2$ to 50~GeV$^2$.
Within the uncertainties of the parameters of the Regge model, a fairly smooth matching can be achieved between the Regge and DIS regions around $Q^2 \approx 2$~GeV$^2$.
Note also that the $F_{3}^{(0)}$ structure function increases to a constant value in the limit $Q^2 \to 0$, given by
    $F_{3}^{(0)}(x, Q^2 \to 0) = A^{p-n} \big[ (W^2-M^2)/\Lambda_R^2 \big]^{\alpha_R}$.
The total Regge region contribution to the $\gw$ box is then found to be
    \mbox{$\Box^{\gw}_{A\, \rm (Reg)} = 0.37(7) \times 10^{-3}$},
where $\approx 40\%$ of the uncertainty comes from the fit parameters, and $\approx 60\%$ is from the systematic model uncertainty.

Following the suggestion by Czarnecki {\it et al.}~\cite{Czarnecki19}, we also considered an extension of the model of Ref.~\cite{Seng19} to include a third vector meson.
However, we found that this did not give any significant improvement in the least squares fit to the data, and no noticeable improvement to the accuracy of $\Box_{A\, \rm (Reg)}^{\gw}$.
A new set of higher precision data in the Regge region would be required in order to further discriminate between the specific models considered here.

\subsection{Resonance and background contributions}
\label{ssec:resonance}

For the resonance region contributions $\Box^{\gw}_{A\, \rm(res)}$ from $W^2 \leq W_0^2$, we consider both explicit resonance states and a nonresonant background on top of which these sit.
The separation of these two contribution to the cross section is generally not unique, so that it necessary to discuss the two components in the same theoretical framework.

\subsubsection{Resonances}

Since only the isoscalar electromagnetic current is relevant for the $\gw$ box correction, isospin 3/2 resonances do not contribute, so that only isospin-1/2 intermediate states need be considered.
Specifically, we include the positive parity Roper resonance $P_{11}(1440)$, and the negative parity $D_{13}(1520)$ and $S_{11}(1535)$ states.

Using the standard notation for the nucleon $\to$ resonance $R$ transition form factors~\cite{Lalakulich05, Lalakulich06, Lalakulich07}, the contribution to the $F_{3 (\rm res)}^{(0)}$ structure function from the spin-3/2 $D_{13}(1520)$ resonance is given by
\begin{eqnarray}
\label{eq:D13res}
F_{3\,(D_{13})}^{(0)}
&=& \frac{2 C_5^A \nu}{3 M M_R} 
    \Big[ C_3^V M \big( 2 M_R (M_R - M) - M \nu + Q^2 \big)      \nonumber\\
& &\qquad\qquad
          +\, M M_R\nu\, (C_4^V + C_5^V) - C_4^V M_R Q^2
    \Big]\, R(W, M_R),
\end{eqnarray}
where $M_R$ denotes the mass of the resonance, $\nu = (M_R^2-M^2+Q^2)/(2 M)$ is the energy transfer in the target rest frame, and $C_{3,4,5}^V$ and $C_5^A$ are the vector and axial vector form factors.
For the spin-1/2 resonances, the structure function contribution is
\begin{eqnarray}
\label{eq:P11S11res}
F_{3\,(P_{11}/S_{11})}^{(0)}
&=& -\frac{F_A \nu}{2M} 
    \Big[ F_1^V Q^2 + 2 F_2^V M (M_R \pm M) \Big]\, R(W, M_R),
\end{eqnarray}
where $F_{1,2}^V$ and $F_A$ are the corresponding vector and axial vector transition form factors, and the $\pm$ signs in the parentheses correspond to the $P_{11}$ and $S_{11}$ states, respectively.
The function $R(W, M_R)$ represents the finite width of the resonance, which we parametrize by a Breit-Wigner shape,
\begin{eqnarray}
R(W, M_R) = \frac{M_R \Gamma_R}{\pi} \frac{1}{(W^2 - M_R^2)^2 + M_R^2 \Gamma_R^2},
\end{eqnarray}
with width $\Gamma_R$.

For the axial vector transition form factors $C_5^A$ and $F_A$ we use the parametrization of Leitner {\it et al.}~\cite{Leitner09}.
For the vector transition form factors, we take the combinations
    $C_i^V = C_i^p + C^n_i$
and $F_i^V = F_i^p + F^n_i$, with the $C_i^{p,n}$ and $F_i^{p,n}$ form factors determined from the MAID2009 electromagnetic helicity amplitudes~\cite{Tiator09}.
In practice, the vector form factors are very similar in both Ref.~\cite{Leitner09} and \cite{Tiator09}.
We also examined the effect of using instead the more recent helicity amplitudes from CLAS at Jefferson Lab~\cite{CLAS19}, but obtained almost identical results.
The relationships between form factors and helicity amplitudes are given in Appendix~B of Ref.~\cite{Leitner09}.

\begin{figure}[t]
\centering
\includegraphics[width=0.75\textwidth]{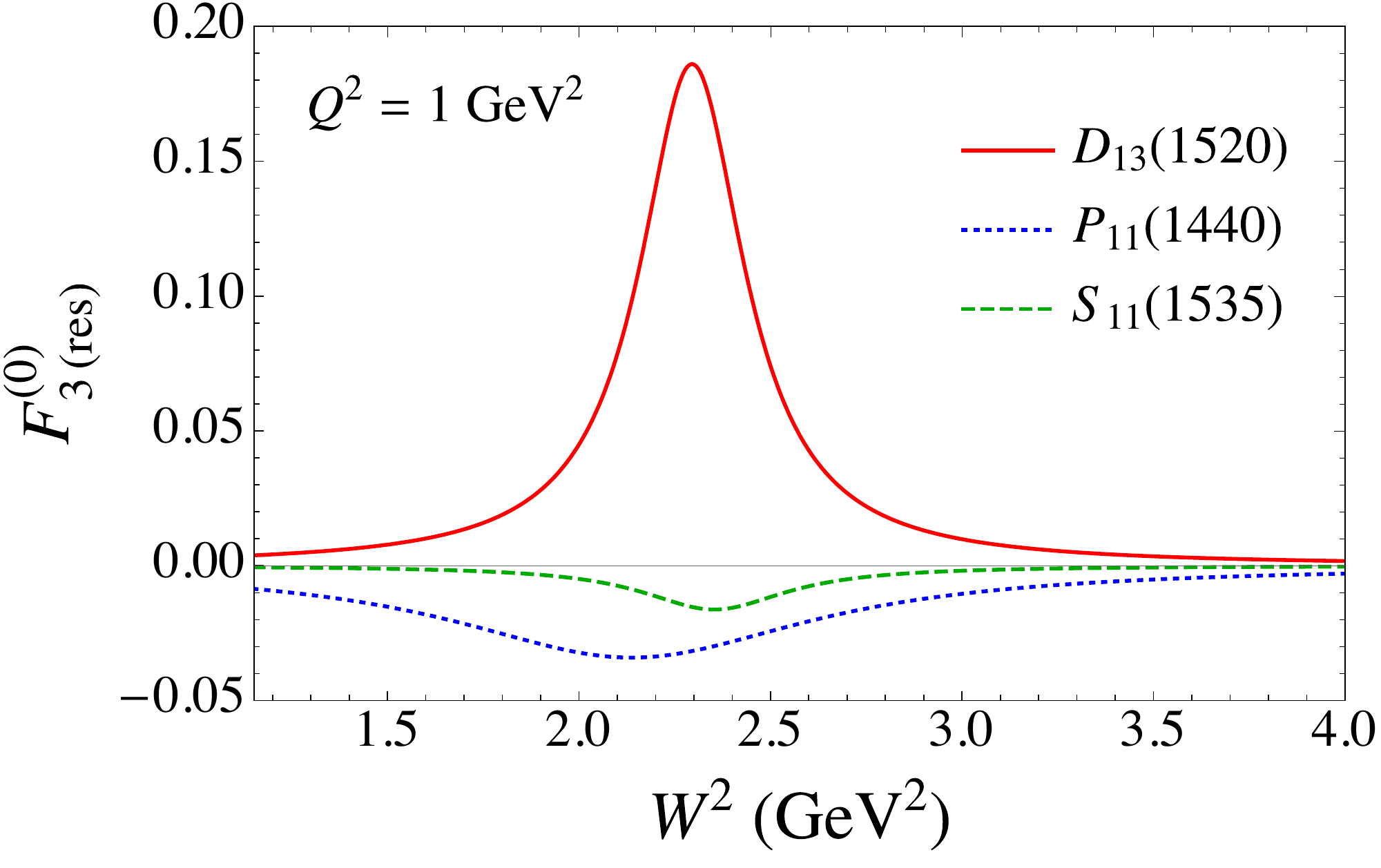}
\caption{Contributions to the $F_{3\, \rm(res)}^{(0)}$ structure function at $Q^2=1~\text{GeV}^2$ from the three dominant $I=1/2$ resonances, including the spin-3/2 $D_{13}(1520)$ (solid red line), and the spin-1/2 $P_{11}(1440)$ (dotted blue line) and 
$S_{11}(1535)$ (dashed green line) resonances, determined from electromagnetic MAID2009 helicity amplitudes~\cite{Tiator09} and axial transition form factors from Leitner {\it et al.}~\cite{Leitner09}.}
\label{fig:f3res}
\end{figure}

The relative sizes of the $P_{11}(1440)$, $D_{13}(1520)$, and $S_{11}(1535)$ resonance contributions to $F_{3 (\rm res)}^{(0)}$ are shown in Fig.~\ref{fig:f3res} at $Q^2=1$~GeV$^2$ as a function of $W^2$.
This illustrates the dominance of the (positive) $D_{13}$ state contribution compared with the somewhat smaller (negative) contributions from the $P_{11}(1440)$ and $S_{11}(1535)$ resonances.
Using the expressions in Eqs.~(\ref{eq:D13res}) and (\ref{eq:P11S11res}) in Eq.~(\ref{eq:boxRe}), we find the contribution to $\Box^{\gw}_A$ from the $D_{13}(1520)$ resonance to be $0.055 \times 10^{-3}$, with the $P_{11}(1440)$ and $S_{11}(1535)$ resonances contributing $-0.008 \times 10^{-3}$ and $-0.003 \times 10^{-3}$, respectively.
Combined, the total resonance contribution is then found to be
    $\Box^{\gw}_{A\, (\rm res)} = 0.04(1)\times 10^{-3}$.

\subsubsection{Nonresonant background}

In addition to the excited state resonances that populate the low-$W$ region, with their prominent peaks and valleys, the physical spectrum also includes contributions from the nonresonant background on top of which the resonances sit.
Of course, it is also clear that any separation of the resonant and nonresonant contributions to the cross section is not unique and necessarily model dependent.
Nevertheless, the background is generally understood to be associated with nonresonant multi-hadron dynamics, which produce a spectrum that is relatively smooth in $W^2$ and $Q^2$, reminiscent of the DIS continuum characterizing the higher-$W$ region, $W^2 > 4$~GeV$^2$, in Fig.~\ref{fig:regions}.

\begin{figure}[t]
\centering
\includegraphics[width=0.75\textwidth]{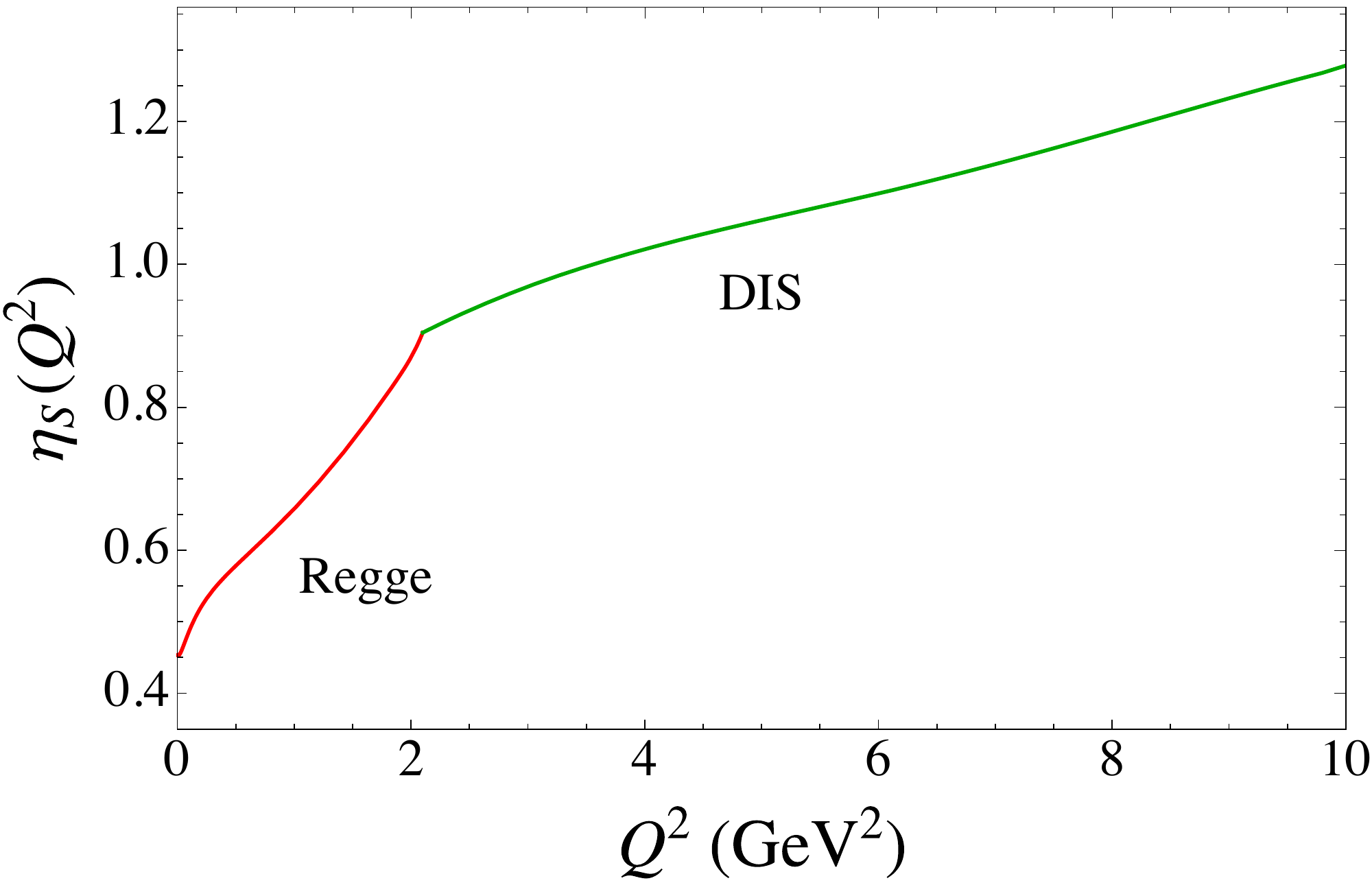}
\caption{Matching function $\eta_S(Q^2)$ versus $Q^2$, which allows the nonresonant background function $F_{3\, (\rm bgd)}^{(0)}$ to match smoothly onto the Regge region $F_{3\, (\rm Reg)}^{(0)}$ (red line) and DIS region $F_{3\, (\rm DIS)}^{(0)}$ (green line) functions at the $W_0^2=4$~GeV$^2$ boundary.}
\label{fig:etaplot}
\end{figure}

Indeed, one approach adopted in the literature (see, {\it e.g.}, Ref.~\cite{Seng18}) has been to extrapolate structure functions from the DIS regime to the $W^2 < W_0^2$ region assuming a smooth functional form for the $x \to 1$ behavior.
However, this introduces an uncontrolled approximation since the DIS function is not constrained by data at such low $W$ values, where incoherent scattering from uncorrelated partons can no longer be considered as a good approximation.
Instead, in our analysis we adapt the Christy-Bosted parametrization~\cite{Bosted10} of the nonresonant (NR) background for the transverse electromagnetic cross section, $\sigma_T^{\rm NR}$.
If we further assume that at fixed $Q^2$ the shape of the $F_3^{(0)}$ background is similar to the shape of the electromagnetic $F_1^\gamma$ structure function background,
    $F_{3\, (\rm bgd)}^{(0)} \propto F^{\gamma}_{1\, (\rm bgd)}$,
then in analogy with Ref.~\cite{Bosted10} the background contribution can be written as
\begin{eqnarray}
F_{3\, (\rm bgd)}^{(0)}(W^2,Q^2)
&=& \eta_S(Q^2) 
\frac{W^2-M^2}{8\pi^2\alpha} 
\bigg( 1 + \frac{W^2-W_\pi^2}{Q^2+\Lambda_{\rm NR}^2} \bigg)^{-1}
\sum_{i=1}^2 
\frac{\sigma_T^{{\rm NR}, i}\, (W-W_\pi)^{i+1/2}}
{(Q^2+a_i^T)^{f_i^T(Q^2)}}.
\label{F30bgd}
\end{eqnarray}
Here, the exponent in the denominator of the sum is 
    $f_i^T(Q^2) \equiv b_i^T+c_i^T Q^2+d_i^T Q^4$,
and the numerical values of the fit parameters
    \{$a_i^T$, $b_i^T$ , $c_i^T$, $d_i^T$, $\Lambda_{\rm NR}^2$, $\sigma_T^{{\rm NR}, i}$\}
are given in Ref.~\cite{Bosted10}.
The overall factor $\eta_S(Q^2)$ is determined by matching $F_{3\, (\rm bgd)}^{(0)}$ along the boundary $W^2=W_0^2$ with the corresponding Regge and DIS region functions, $F_{3\, (\rm Reg)}^{(0)}$ and $F_{3\, (\rm DIS)}^{(0)}$, along the $Q^2 < Q_0^2$ and $Q^2 > Q_0^2$ boundaries, respectively.
The matching function $\eta_S(Q^2)$ is illustrated in Fig.~\ref{fig:etaplot} as a function of $Q^2$, for the Regge and DIS regions.
At intermediate $Q^2$ values, $Q^2 \approx 2-8$~GeV$^2$, the matching function is within $\approx 10\%$ of unity, but decreases more steeply as $Q^2 \to 0$.  The parameterization in (\ref{F30bgd}) only admits a central value, and thus alone, it cannot provide one with an uncertainty estimate.  In order to estimate an uncertainty to $F_{3\text{(bgd)}}^{(0)}$, we compare (\ref{F30bgd}) to the extrapolated $F_{3\, (\rm DIS)}^{(0)}$ over the region $W^2\leq W_0^2$ and find their difference is on the order of $5\%$.

\begin{figure}[t]
\centering
\includegraphics[width=0.65\textwidth]{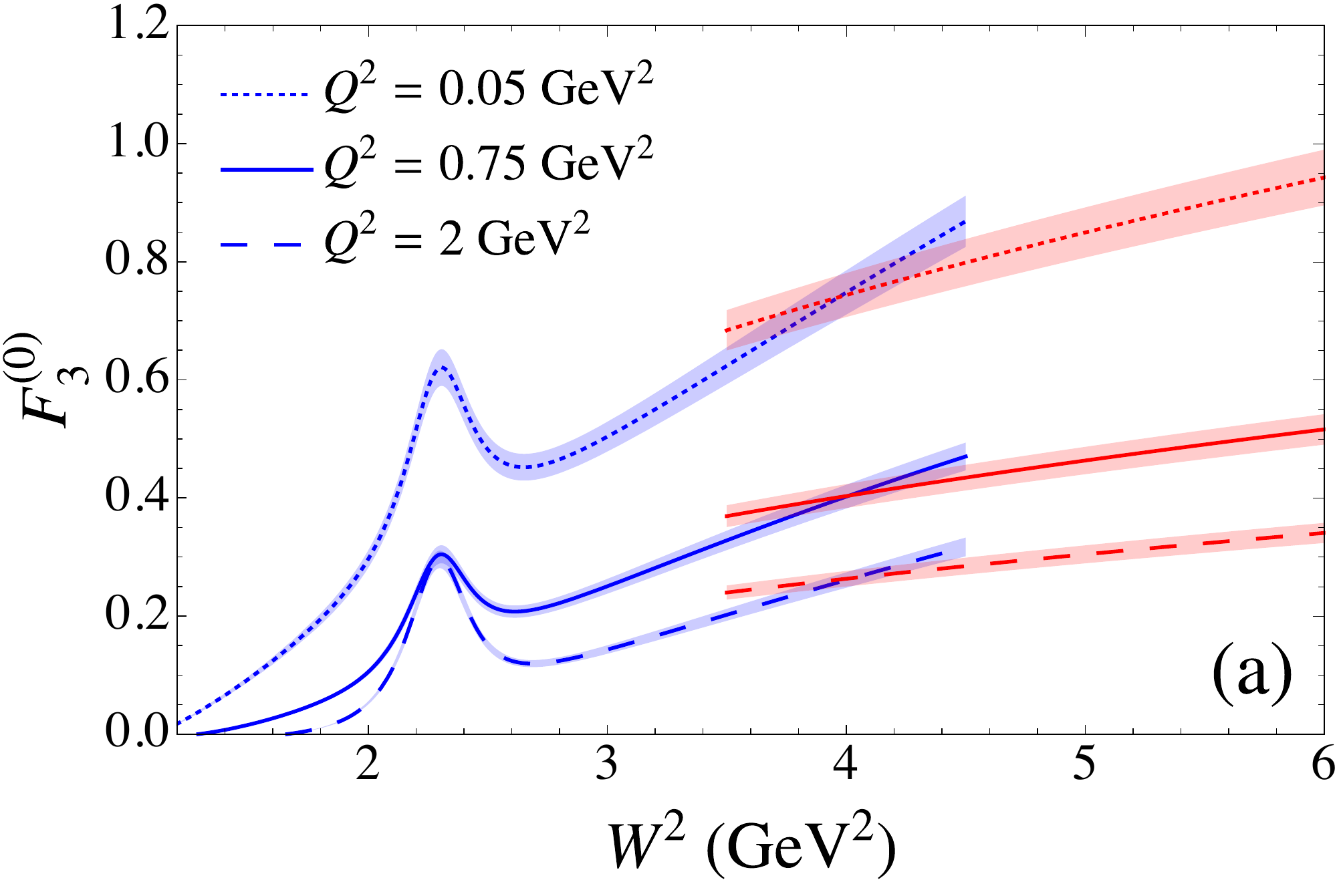}
\includegraphics[width=0.65\textwidth]{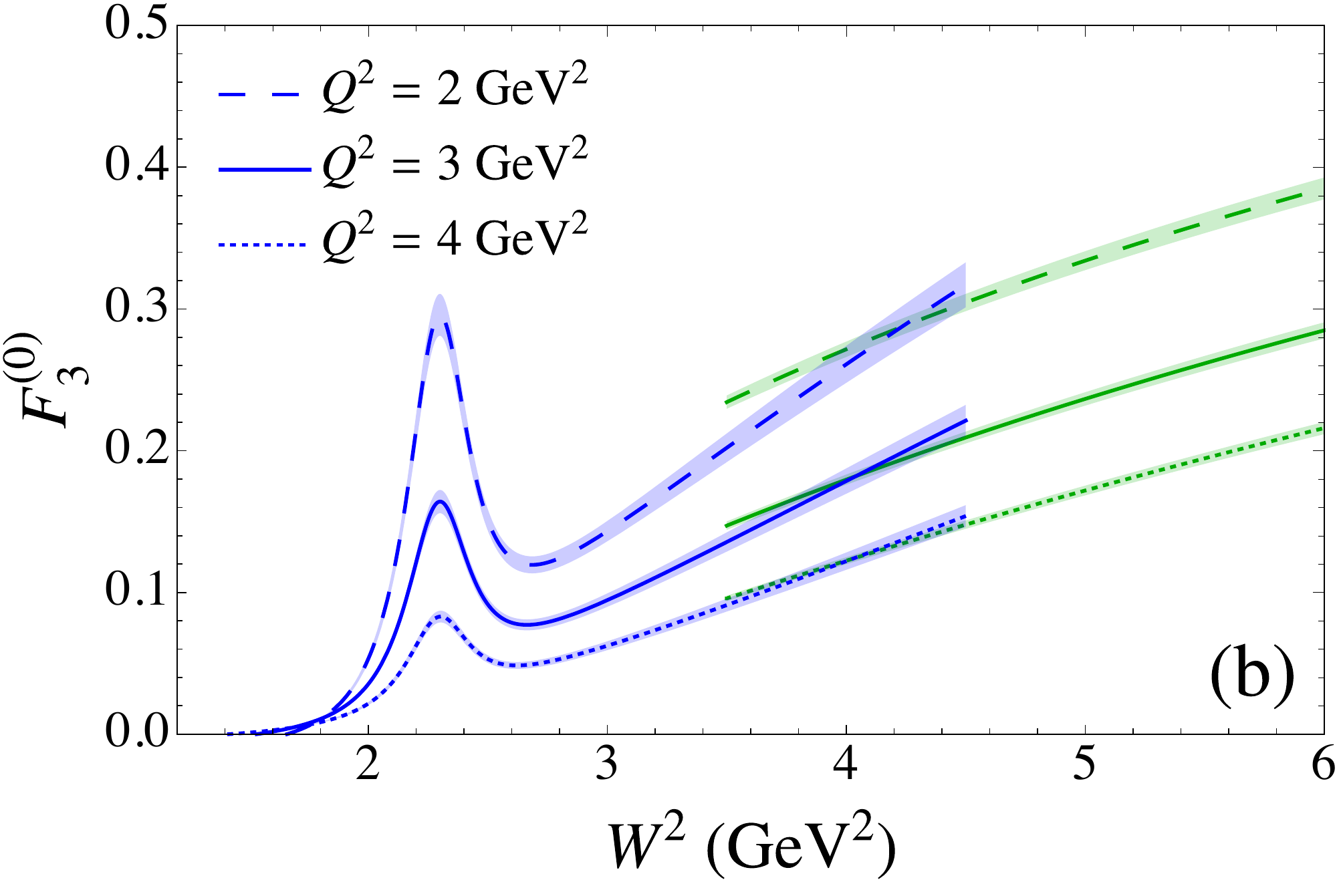}
\caption{{\bf (a)} Matching of the $F_{3}^{(0)}$ structure function in the resonance $+$ background region (blue lines and bands) versus $W^2$ with
the Regge region structure function $F_{3\, (\rm Reg)}^{(0)}$ at fixed $Q^2=0.05$~GeV$^2$ (dotted lines), 0.75~GeV$^2$ (solid lines), and 2~GeV$^2$ (dashed lines).
{\bf (b)} Matching to the DIS region structure function $F_{3\, (\rm DIS)}^{(0)}$ at fixed $Q^2=2$~GeV$^2$ (dashed lines), 3~GeV$^2$ (solid lines), and 4~GeV$^2$ (dotted lines).}
\label{fig:matching_res_DIS_Reg}
\end{figure}

The resulting matching of the $F_{3}^{(0)}$ structure function between the resonance $+$ background region (iv) and the DIS region (ii) and Regge region (iii) is illustrated in Fig.~\ref{fig:matching_res_DIS_Reg} as a function of $W^2$ for various fixed values of $Q^2$, from $Q^2=0.05$~GeV$^2$ to $Q^2=4$~GeV$^2$.
Beyond $Q^2 \approx 10$~GeV$^2$ the resonance contributions become very small.
The matching of the structure functions at $W^2=4$~GeV$^2$ suggests a relatively smooth transition from the resonance region at low $W$ to the higher-$W$ regions described by the Regge and DIS parametrizations at low and high $Q^2$ values, respectively.

Integrating the $F_{3 \rm (bgd)}^{(0)}$ structure function (\ref{F30bgd}) into Eq.~(\ref{eq:boxRe}), we find the total nonresonant background contribution the $\gw$ box correction is 
    \mbox{$\Box^{\gw}_{A\, \rm (bgd)} = 0.15(1) \times 10^{-3}$},
of which $\approx 95\%$ comes from the region $Q^2 > 2$~GeV$^2$.
For the combined resonance and background contributions, where the former is dominated by the $D_{13}$ state, integrating over the sum 
    $F_{3\, \rm (res)}^{(0)} + F_{3\, \rm (bgd)}^{(0)}$
in (\ref{eq:boxRe}) yields the total $W_\pi^2 \le W^2 \le 4$~GeV$^2$ region contribution to the $\gw$ box correction of $0.19(2) \times 10^{-3}$.

\section{Impact of the $\gw$ box on $V_{ud}$ determination}
\label{sec:results}

Putting together the contributions from the various regions in Fig.~\ref{fig:regions}, the total $\Box_A^{\gw}$ correction from the elastic, resonance, DIS and Regge regions is summarized in Table~\ref{tab:TableI}.
For our nominal boundary between the DIS and Regge regions of $Q_0^2=2$~GeV$^2$, we find the total box correction to be $\Box_A^{\gw} = 3.90(9) \times 10^{-3}$, of which $\approx 59\%$ comes from DIS and $\approx 27\%$ from elastic intermediate states.
The effect of lowering the DIS--Regge boundary to $Q_0^2=1$~GeV$^2$ is to increase the DIS contribution by $\approx 3\%-4\%$ of the total $\Box_A^{\gw}$, with an almost identical compensating decrease in the Regge component, rendering the sum almost unchanged.
This provides confidence in the robustness of our calculation of the total $\Box_A^{\gw}$ correction, which has relatively weak dependence on how the individual pieces are computed.

\begin{table}[t]
\centering
\caption{Summary of corrections to $\Box_A^{\gw}$ (in units of $10^{-3}$) from the various kinematic regions in Fig.~\ref{fig:regions}. To compare the results of the present analysis (SBM) with previous results from SGRM~\cite{Seng19} and CMS~\cite{Czarnecki19}, the $Q^2 > Q_0^2$ part of the nonresonant background is combined with the DIS contribution, and the $Q^2 < Q_0^2$ part of the background is combined with the Regge contribution. The DIS contributions marked with an asterisk (*) have been calculated with $\alpha=\alpha(0)$. The final two rows give results for $\Delta_R^V$ and $|V_{ud}|^2$, and our central results are highlighted in boldface.\\}
\setlength\tabcolsep{6pt}
\begin{ruledtabular}
\begin{tabular}{l c c c c}
\bm{$\Box_A^{\gw}~(\times 10^{-3})$} 
& \multicolumn{2}{c}{SBM} & {SGRM~\cite{Seng19}} & {CMS~\cite{Czarnecki19}} \\
\cline{2-3}
~~~~~$Q_0^2$~(GeV$^2$) & {\bf 2.0} & {1.0} & {2.0}   & {1.1}  \\ \hline
elastic
& 1.05(4) &  1.05(4) & 1.06(6) & 0.99(10) \\ 
resonance
& 0.04(1) &  0.04(1) & {---} & {---}\\
DIS + ($Q^2 > Q_0^2$) bgd 
& 2.29(3) &  2.43(5)  & *2.17(0)~ & *2.29(2)~ \\
Regge + ($Q^2 < Q_0^2$) bgd
& 0.52(7) & 0.39(5) & 0.56(8) & 0.25(2) \\
\cline{2-5}
~~~~~Total
& {\bf 3.90(9)} & 3.91(9) & ~3.79(10) & ~3.52(11) \\
\hline
$\Delta_R^V$ & {\bf 0.02472(18)} & 0.02474(18) & 0.02467(22) & 0.02426(32)\\
$|V_{ud}|^2$ & {\bf 0.94805(26)} & 0.94803(26) & 0.94809(28) & 0.94847(35)\\
\end{tabular}
\end{ruledtabular}
\label{tab:TableI}
\end{table}

Our total $\gw$ correction is $\approx 10$\% larger than the Czarnecki {\it et al.} (CMS)~\cite{Czarnecki19} result, and $\approx 3$\% larger than the Seng {\it et al.} (SGRM)~\cite{Seng19} value.
To identify the sources of the differences, we note that the boundaries between the various regions in Refs.~\cite{Czarnecki19, Seng19} are not identical to those in Fig.~\ref{fig:regions}.
In particular, SGRM integrate the Regge and DIS contributions down to the inelastic threshold.
To make a more direct comparison, we therefore add in Table~\ref{tab:TableI} the low-$Q^2$ ($Q^2 < Q_0^2$) part of the nonresonant background to the Regge contribution, and add the high-$Q^2$ ($Q^2 > Q_0^2$) part of the background to the DIS component.
Compared with the SGRM result~\cite{Seng19}, which also uses $Q_0^2=2$~GeV$^2$ for the Regge--DIS boundary, our $Q^2 > Q_0^2$ contribution is larger by $0.12 \times 10^{-3}$, which accounts for virtually all of the difference in the totals.
SGRM~\cite{Seng19} also isolate a ``single pion'' contribution, which is calculated separately using chiral perturbation theory, whereas in our analysis this is effectively accounted for in the nonresonant background.
Finally, SGRM compute the DIS contributions with a constant $\alpha = \alpha(0)$, however, if the running of $\alpha(Q^2)$ is taken into account via Eq.~(\ref{eq:F3DIS}), their result agrees well with ours. 

By contrast, the CMS~\cite{Czarnecki19} analysis begins their perturbative threshold at $Q_0^2=1.1$~GeV$^2$, so that the relative contributions from the low-$Q^2$ and high-$Q^2$ regions can be more directly compared with our results for $Q_0^2=1$~GeV$^2$ in Table~\ref{tab:TableI}.
Both the $Q^2 < Q_0^2$ and $Q^2 > Q_0^2$ contributions from CMS are smaller than those in our analysis, especially the low-$Q^2$ result, leading to the $\approx 10$\% smaller value in Ref.~\cite{Czarnecki19} for the total $\Box_A^{\gw}$ correction.
An additional complication of the comparison between the different calculations is that the elastic contribution in Ref.~\cite{Czarnecki19} is only integrated over a limited range of $Q^2$, whereas in Ref.~\cite{Seng19} and in our analysis the elastic contribution is integrated over all $Q^2$. 
We will return to this point later in this section.
What can be seen from the results in Table~\ref{tab:TableI}, however, is that the value of $\Box_A^{\gw}$ has increased substantially compared to the previously adopted value of $3.26(19) \times 10^{-3}$ from Marciano and Sirlin~\cite{Marciano06}, but the uncertainty on this correction has decreased by a factor $\sim 2$.

\begin{figure}[t]
\centering
\includegraphics[width=0.95\textwidth]{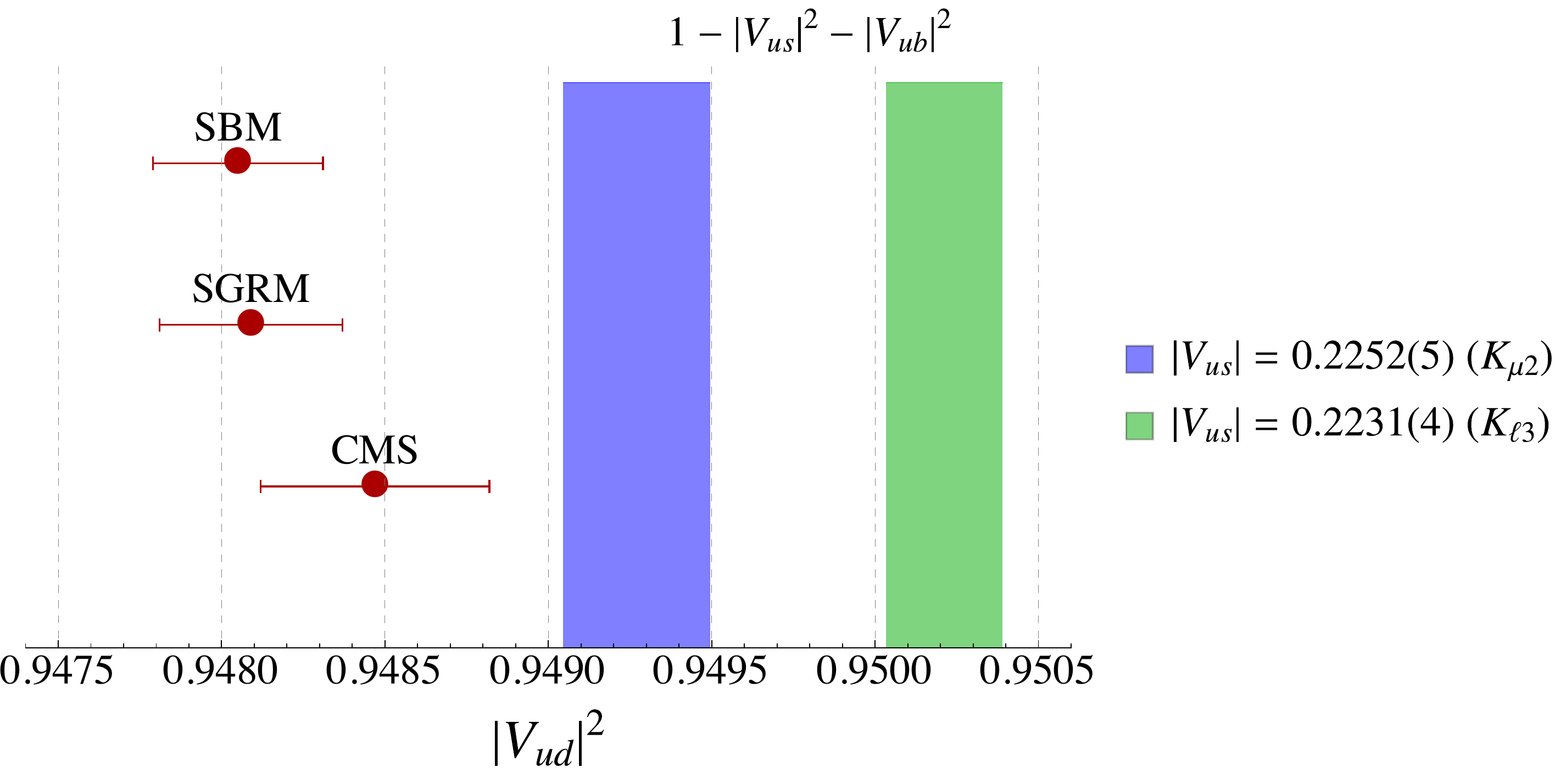}
\caption{Comparison of the $|V_{ud}|^2$ matrix element from the present work (SBM) with values derived from the analyses of Refs.~\cite{Seng19} (SGRM) and \cite{Czarnecki19} (CMS). The vertical blue and green bands denote the values of $1-|V_{us}|^2 - |V_{ub}|^2$ obtained using data from $K_{\mu 2}$ and $K_{\ell 3}$ decays, respectively, together with $|V_{ub}|=0.0038(2)$~\cite{PDG}. The gap between our value of $|V_{ud}|^2$ and the bands suggests a violation of CKM unitarity at the (3--5)$\sigma$ level.}
\label{fig:Vud}
\end{figure}

The impact of our calculated $\gw$ correction on the CKM matrix element $V_{ud}$ can be quantified by noting the relation between $\Box_A^{\gw}$ and the total inner radiative correction $\Delta_R^V$ in the denominator of Eq.~(\ref{eq:Vud}).
To lowest order~\cite{Sirlin78,Towner92},
\begin{subequations}
\label{eq:delRV}
\begin{eqnarray}
\label{eq:delRV1}
\Delta_R^V
&=& \frac{\alpha}{2\pi}
    \bigg[ 3 \ln\frac{M_W}{M} - 4 \ln c_W \bigg] + 2\, \Box_A^{\gw} \\
&=& \frac{\alpha}{2\pi}
    \bigg[ 3 \ln\frac{M_Z}{M} -   \ln c_W \bigg] + 2\, \Box_A^{\gw},
\label{eq:delRV2}
\end{eqnarray}
\end{subequations}
where $c_W = M_W/M_Z$, and $\alpha$ is taken at the Thomson limit, $Q^2=0$.
The leading $\ln M_Z/M$ term in Eq.~(\ref{eq:delRV2}) can then be resummed with the aid of a renormalization group analysis, as in Ref.~\cite{Marciano86}, with the resulting replacement
\begin{eqnarray}
\label{eq:resummed}
1 + \frac{2\alpha}{\pi} \ln \frac{M_Z}{M} &\to & 1.02248.
\end{eqnarray}
Incorporating the resummed result (\ref{eq:resummed}) in Eq.~(\ref{eq:delRV2}), together with other small corrections, leads to the simple relationship
\begin{equation}\label{eq:delRV3}
\Delta_R^V = 0.01691 + 2\, \Box_A^{\gw}.    
\end{equation}
Using our preferred value for the total $\Box_A^{\gw}=3.90(9)\times 10^{-3}$ in
Eq.~(\ref{eq:delRV3}), together with Eq.~(\ref{eq:Vud}), we obtain our best estimate for the $V_{ud}$ CKM matrix element,
\begin{equation}
    |V_{ud}|^2 = 0.94805(26).
\end{equation}
This result is approximately $4\sigma$ below the expected value based on the unitarity prediction $|V_{ud}|^2+|V_{us}|^2+|V_{ub}|^2=1$.
A comparison of this result to CMS and SGRM is given in Table~\ref{tab:TableI}, and is depicted graphically in Fig.~\ref{fig:Vud}.
We note that Refs.~\cite{Czarnecki19, Seng19} take a different approach to accounting for higher order effects than we have, resulting in slightly different expressions for $\Delta_R^V$ than that given in Eq.~(\ref{eq:delRV3}). In particular, CMS (see Eq.~(19) of \cite{Czarnecki19}) apply enhancement factors of 1.022 and 1.065 to the contributions in Table~\ref{tab:TableI} with $Q^2$ below and above $Q_0^2$, respectively.

\section{Conclusions}
\label{sec:conclusion}

In this work we have presented a comprehensive new analysis of the $\gw$ interference contribution to the neutron \mbox{$\beta$-decay} matrix element using a dispersive relations framework that has previously been applied successfully for $\gz$ box corrections.
The evaluation of the $\gw$ correction relies on knowledge of the isoscalar part of the parity-odd $\gw$ structure function, $F_3^{(0)}$, which is not directly accessible experimentally, but can be modelled from existing phenomenology.

Separating the $F_3^{(0)}$ structure function into distinct contributions from the nucleon elastic, resonance, deep-inelastic, and Regge regions, we have used the latest phenomenological and theoretical constraints to compute the contributions from each region and estimate the associated uncertainties.
We find the real part of the box correction to be \mbox{$\Re \Box^{\gw}_A = 3.90(9) \times 10^{-3}$}, with the main contribution ($\sim 60\%$) coming from the high-$Q^2$, high-$Q^2$ DIS region, with the next largest contribution ($\sim 30\%$) coming from elastic intermediate states.
The largest uncertainty comes from the Regge region, which we estimate is $\sim 15\%$, followed by the nucleon elastic contribution.
While the detailed breakdown into the various components depends on the choices of the boundaries between the regions, the total correction is fairly robust, and only weakly dependent on the decomposition.

The new, improved correction is larger than that found in recent analyses~\cite{Marciano06, Czarnecki19, Seng18, Seng19} (though consistent within uncertainties with the SGRM result~\cite{Seng19}).
The resulting value of the CKM matrix element $|V_{ud}|^2 = 0.94805(26)$ extracted via superallowed $\beta$ decays represents a 4$\sigma$ violation of unitarity.

It has been suggested recently~\cite{Seng:2019plg} that the $\gw$ box diagram can be computed in lattice QCD using the Feynman-Hellmann theorem.
Direct calculation is rather challenging, and the simpler $\gw$ correction to the rate of the semileptonic pion decay was calculated in Ref.~\cite{Feng:2020zdc}.
Using a combination of dispersion theory and phenomenology, the results of this lattice study of the pion were converted to the more relevant case of the neutron~\cite{Seng:2020wjq}, confirming the top-row CKM unitarity deficit.
Future lattice calculations of semileptonic baryon decays would be able to test the unitarity violations explicitly.
Although difficult to determine empirically, the $F_3^{\gw}$ structure function could also be inferred from model-independent relations between $\gw$ and $\gz$ interference structure functions, as well as input from neutrino and antineutrino inclusive DIS.

Finally, we reiterate the observation of Hardy and Towner~\cite{Hardy15} that improvements to $\Delta_R^V$ are the highest priority in any real improvement in the unitarity test from $0^+\to 0^+$ $\beta$ decay.
We note that there is significant variation in the literature~\cite{Towner92, Towner08, Czarnecki19, Seng19} in the treatment of effects beyond leading order in $\alpha$ in the $\gw$ box, as well as other contributions to $\Delta_R^V$.
This suggests that such higher order effects are worthy of further theoretical consideration.

\section*{Acknowledgments}

We acknowledge helpful discussions with C.-Y.~Seng, M.~Gorchtein, M.~Ramsey-Musolf, H.~Patel, C.~McRae, and J.~Qiu.
We also thank the organizers of the 2018 program ``Bridging the Standard Model to New Physics with the Parity Violation Program at MESA'', hosted by MITP Mainz.
This work was supported by the Natural Sciences and Engineering Research Council (Canada), and the US Department of Energy contract DE-AC05-06OR23177, under which Jefferson Science Associates, LLC operates Jefferson Lab.

\newpage
\appendix
\section{Relations between $\gw$, $\gz$ and $W^\pm$ DIS structure functions}
\label{app:F3DIS}

In this appendix we derive relations between parity-violating leading twist structure functions for charged current neutrino and antineutrino scattering, $\gz$ interference, and the nondiagonal $\gw$ case relevant for the present analysis of the $\gw$ box correction.
For simplicity, we consider the structure functions at lowest order in $\alpha_s$, which allows them to be expressed entirely in terms of PDFs.
Inclusion of higher order corrections is straightforward, and involves generalizing the relations by convolutions between the PDFs and hard scattering cross sections.

To relate the flavor nondiagonal structure functions to the flavour diagonal structure functions that appear in inclusive DIS processes, we consider the transition $n \to p$ matrix element of nonlocal, leading-twist light-cone operators separated by a light-like distance $z$.
Following the discussion of Mankiewicz~{\it et al.}~\cite{Mankiewicz98}, at forward angles, this can be written as
\begin{equation}
\langle p(P)|\widehat{\cal O}_{ud}(0,z)|n(P)\rangle\Big|_{z^2=0}\
\sim\ \int_0^1 dx\, 
\left[ e^{-i x (P \cdot z)} f^{ud}(x)
   \pm e^{ i x (P \cdot z)}\bar{f}^{ud}(x)
\right],
\end{equation}
where $\langle p(P) |$  and $|n(P) \rangle$ represent proton and neutron states, respectively, with momentum~$P$, and $f^{ud}$ and $\bar{f}^{ud}$ denote flavor nondiagonal quark and antiquark distributions, respectively, for $n \to p$ transitions, evaluated at parton momentum fraction $x$.
The sign in front of the $\bar{f}^{ud}$ distribution is determined by the charge conjugation properties of the leading twist operator in the local limit.
The quark bilinear operator $\widehat{\cal O}_{qq'}(0,z)$ is defined as a product of two quark fields of flavor $q$ and $q'$, separated by a light-like distance $z$,
\begin{eqnarray}
\widehat{\cal O}_{qq'}(0,z) = \overline{\psi}_q(0)\, \Gamma\, \psi_{q'}(z) \Big|_{z^2=0},
\end{eqnarray}
for a given Dirac operator $\Gamma$.
As discussed by Mankiewicz {\it et al.}~\cite{Mankiewicz98} , isospin symmetry relations relate the flavor nondiagonal matrix elements to flavor diagonal ones via
\begin{eqnarray}
\begin{aligned}
\label{eq.O_relations}
\langle p|\widehat{\cal O}_{ud}|n\rangle
&= \langle p|\widehat{\cal O}_{uu}|p\rangle
 - \langle p|\widehat{\cal O}_{dd}|p\rangle    \\
&= \langle n|\widehat{\cal O}_{dd}|n\rangle
 - \langle n|\widehat{\cal O}_{uu}|n\rangle.
\end{aligned}
\end{eqnarray}
in the proton and neutron, respectively.
The relations (\ref{eq.O_relations}) then imply simple relations between the flavor nondiagonal and flavor diagonal distribution functions, and hence between the $F_3^\gw$ and $F_3^\gz$ structure functions.
In particular, for the isoscalar electromagnetic current contribution to the $F_3^\gw$ structure function, from Eq.~(\ref{eq.O_relations}) we can write
\begin{eqnarray}
\label{eq.F30}
F_3^{(0)}(x)
&=& (e_u+e_d) 
    \big( f^{ud}(x) - \bar{f}^{ud}(x)
    \big)
\nonumber\\
&=& (e_u+e_d)
\Big[
    \big( f^{uu}(x) - f^{dd}(x)
    \big)
  - \big( \bar{f}^{uu}(x) - \bar{f}^{dd}(x)
    \big)
\Big]
\nonumber\\
&=& (e_u+e_d)
\Big[
    \big( u(x)-\bar{u}(x) \big) - \big( d(x)-\bar{d}(x) \big)
\Big],
\end{eqnarray}
where we have used the relations
$f^{qq}(x) \equiv q(x)$ and 
$\bar{f}^{qq}(x) \equiv \bar q(x)$,
together with isospin symmetry for PDFs in the proton and neutron,
    $u_n = d_p \equiv d$
and $d_n = u_p \equiv u$.

For the isovector electromagnetic current contribution to $F_3^\gw$, we have the combination
\begin{eqnarray}
F_3^{(1)}(x)
&=& (e_u-e_d) 
    \big( f^{ud}(x)+\bar{f}^{ud}(x) 
    \big)
\nonumber\\
&=& (e_u-e_d) 
\Big[
    \big(f^{uu}(x) - f^{dd}(x)
    \big)
 + \big(\bar{f}^{uu}(x) - \bar{f}^{dd}(x)
   \big)
\Big]
\nonumber\\
&=& (e_u-e_d) 
\Big[
    \big(u(x) + \bar{u}(x)\big) - \big(d(x) + \bar{d}(x)\big)
\Big].
\end{eqnarray}
By crossing symmetry, the structure function $F_3^{(1)}$ does not contribute to $\Re \Box^\gw_A(E=0)$.
Interestingly, the $F_3^{(1)}$ structure function is related to the Gottfried sum rule  integrand~\cite{GSR}, which involves the isovector electromagnetic $F_1$ (or $F_2$) structure function,
\begin{eqnarray}
F_1^{\gamma p}(x)- F_1^{\gamma n}(x)
&=& (e_u^2 - e_d^2)
    \Big[ \big(u(x) + \bar{u}(x)\big) - \big(d(x) + \bar{d}(x)\big)
    \Big]  
\nonumber\\
&=& (e_u + e_d)\, F_3^{(1)}(x),
\end{eqnarray}
which has been studied experimentally in inclusive charged lepton DIS from protons and deuterons~\cite{Accardi:2019ofk}.

For the parity-violating $\gz$ interference structure function
    $F_3^\gz = \sum_q 2\, e_q\, g_A^q\, (q-\bar{q})$~\cite{PDG},
where $g_A^q = \pm \frac12$ for $u$- and $d$-type quarks, respectively, we have explicitly for proton and neutron targets,
\begin{subequations}
\begin{eqnarray}
F_{3p}^\gz(x)
&=& e_u \big( u(x)-\bar{u}(x) \big)
 -  e_d \big( d(x)-\bar{d}(x) \big)
 -  e_s \big( s(x)-\bar{s}(x) \big)
 +  e_c \big( c(x)-\bar{c}(x) \big),~~~~~~        \\
F_{3n}^\gz(x)
&=& e_u \big( d(x)-\bar{d}(x) \big)
 -  e_d \big( u(x)-\bar{u}(x) \big)
 -  e_s \big( s(x)-\bar{s}(x) \big)
 +  e_c \big( c(x)-\bar{c}(x) \big),~~~~~~
\end{eqnarray}
\end{subequations}
with $e_s=e_d=-1/3$ and $e_c=e_u=2/3$.
Taking sums and differences of the proton and neutron functions, one has
\begin{subequations}
\label{eq.F3gz}
\begin{eqnarray}
F_{3p}^\gz(x) - F_{3n}^\gz(x)
&=& (e_u+e_d)
\Big[ 
    \big( u(x)-\bar{u}(x) \big) - \big( d(x)-\bar{d}(x) \big)
\Big],
\label{eq.F3gzminus}\\
F_{3p}^\gz(x) + F_{3n}^\gz(x)
&=& (e_u-e_d)
\Big[
    \big( u(x)-\bar{u}(x) \big) + \big( d(x)-\bar{d}(x) \big)
\Big]
\nonumber\\
&-& 2 e_s \big( s(x)-\bar{s}(x) \big)
 +  2 e_c \big( c(x)-\bar{c}(x) \big)
\nonumber\\
& & \hspace*{-2cm}
 =  \big( u(x)-\bar{u}(x) \big) 
 +  \big( d(x)-\bar{d}(x) \big)
 +  \tfrac23 \big( s(x)-\bar{s}(x) \big)
 +  \tfrac43 \big( c(x)-\bar{c}(x) \big).~~~~
\label{eq.F3gzplus}
\end{eqnarray}
\end{subequations}

For charged-current DIS, the parity-violating inclusive $\nu p$ and $\bar\nu p$ structure functions, denoted by $F_{3p}^{W^+}$ and $F_{3p}^{W^-}$, respectively, provide additional combinations of PDFs not accessible with neutral currents.
Taking the sum of the $\nu p$ and $\bar\nu p$ (or, in fact, $\nu n$ and $\bar\nu n$) structure functions, we have
\begin{eqnarray}
\label{eq.F3W}
F_3^W(x)
&\equiv& \frac12 \Big[ F_{3p}^{W^+}\!(x) + F_{3p}^{W^-}\!(x) \Big]
 =  \frac12 \Big[ F_{3n}^{W^-}\!(x) + F_{3n}^{W^+}\!(x) \Big]
\nonumber\\
&=& \big( u(x)-\bar{u}(x) \big)
 +  \big( d(x)-\bar{d}(x) \big) 
 +  \big( s(x)-\bar{s}(x) \big)
 +  \big( c(x)-\bar{c}(x) \big).
\end{eqnarray}
Comparing Eq.~(\ref{eq.F3W}) with Eqs.~(\ref{eq.F3gz}) and Eq.~(\ref{eq.F30}), we can write the relations between the $\gw$, $\gz$ and $W^\pm$ structure functions as
\begin{subequations}
\begin{eqnarray}
F_3^{(0)}(x) &=& F_{3p}^\gz(x) - F_{3n}^\gz(x),
\label{eq.F30gz}
\\
F_3^W(x) &\approx& F_{3p}^\gz(x) + F_{3n}^\gz(x),
\label{eq.F3W_F3gZ}
\end{eqnarray}
\end{subequations}
where the first result holds in general, while the second result only holds if one ignores strange and charm contributions.

Finally, we compare the above relations with the results obtained by Seng~{\it et al.} (SGRM)~\cite{Seng19}, who, after adjusting for a different normalization from ours, give the relation
\begin{equation}
F_{3, \mbox{\tiny SGRM}}^{(0)}
= \frac12 (e_u+e_d) \big(u-\bar{d}\big)
= \frac12 (e_u+e_d)
  \Big[ \big(u-\bar{u}\big) + \big(\bar{u}-\bar{d}\big)
  \Big].
\label{eq.F30Chen}
\end{equation}
This is in disagreement with the result in Eq.~(\ref{eq.F30}) and with the relation in Eq.~(\ref{eq.F30gz}).
However, this does not affect their numerical results for two reasons.
First, they assume a symmetric sea, $\bar{u}=\bar{d}$, and second, their formulation makes use of the integral $\int_0^1 dx F_3^{(0)}(x)$.
When combined with the first assumption, this integrates to $(e_u+e_d)$, which is the same as would be obtained by integrating Eq.~(\ref{eq.F30}). 
On the other hand, the assumption that $\bar{u}=\bar{d}$ is not consistent with global QCD analyses of high-energy scattering data, which generally find an enhancement of $\bar d$ over $\bar u$ at values of $x \gtrsim 0.01$, and give integrated values for the lowest moment of
    $\int_0^1 dx\, (\bar{u}-\bar{d}) \sim -0.1$.
The values of the moment have large uncertainty, however, due to assumptions about extrapolation into the unmeasured region at low $x$~\cite{CJ15, MMHT14, ABKM09, JAM19, CT18} and it is not even clear whether the integral down to $x=0$ converges~\cite{NNPDF}.

\newpage

\end{document}